\message
{JNL.TEX version 0.95 as of 5/13/90.  Using CM fonts.}

\catcode`@=11
\expandafter\ifx\csname inp@t\endcsname\relax\let\inp@t=\input
\def\input#1 {\expandafter\ifx\csname #1IsLoaded\endcsname\relax
\inp@t#1%
\expandafter\def\csname #1IsLoaded\endcsname{(#1 was previously loaded)}
\else\message{\csname #1IsLoaded\endcsname}\fi}\fi
\catcode`@=12

\font\twelverm=cmr12			\font\twelvei=cmmi12
\font\twelvesy=cmsy10 scaled 1200	\font\twelveex=cmex10 scaled 1200
\font\twelvebf=cmbx12			\font\twelvesl=cmsl12
\font\twelvett=cmtt12			\font\twelveit=cmti12
\font\twelvesc=cmcsc10 scaled 1200	\font\twelvesf=cmss12
\font\twelvemib=cmr10 scaled 1200
                     
\font\tenmib=cmr10
\font\eightmib=cmr10 scaled 800 

\skewchar\twelvei='177			\skewchar\twelvesy='60
\skewchar\twelvemib='177

\newfam\mibfam

\def\twelvepoint{\normalbaselineskip=12.4pt plus 0.1pt minus 0.1pt
  \abovedisplayskip 12.4pt plus 3pt minus 9pt
  \belowdisplayskip 12.4pt plus 3pt minus 9pt
  \abovedisplayshortskip 0pt plus 3pt
  \belowdisplayshortskip 7.2pt plus 3pt minus 4pt
  \smallskipamount=3.6pt plus1.2pt minus1.2pt
  \medskipamount=7.2pt plus2.4pt minus2.4pt
  \bigskipamount=14.4pt plus4.8pt minus4.8pt
  \def\rm{\fam0\twelverm}          \def\it{\fam\itfam\twelveit}%
  \def\sl{\fam\slfam\twelvesl}     \def\bf{\fam\bffam\twelvebf}%
  \def\mit{\fam 1}                 \def\cal{\fam 2}%
  \def\sc{\twelvesc}		   \def\tt{\twelvett}%
  \def\sf{\twelvesf}               \def\mib{\fam\mibfam\twelvemib}%
  \textfont0=\twelverm   \scriptfont0=\tenrm   \scriptscriptfont0=\sevenrm
  \textfont1=\twelvei    \scriptfont1=\teni    \scriptscriptfont1=\seveni
  \textfont2=\twelvesy   \scriptfont2=\tensy   \scriptscriptfont2=\sevensy
  \textfont3=\twelveex   \scriptfont3=\twelveex\scriptscriptfont3=\twelveex
  \textfont\itfam=\twelveit
  \textfont\slfam=\twelvesl
  \textfont\bffam=\twelvebf \scriptfont\bffam=\tenbf
                            \scriptscriptfont\bffam=\sevenbf
  \textfont\mibfam=\twelvemib \scriptfont\mibfam=\tenmib
                              \scriptscriptfont\mibfam=\eightmib
  \normalbaselines\rm}


\mathchardef\alpha="710B
\mathchardef\beta="710C
\mathchardef\gamma="710D
\mathchardef\delta="710E
\mathchardef\epsilon="710F
\mathchardef\zeta="7110
\mathchardef\eta="7111
\mathchardef\theta="7112
\mathchardef\iota="7113
\mathchardef\kappa="7114
\mathchardef\lambda="7115
\mathchardef\mu="7116
\mathchardef\nu="7117
\mathchardef\xi="7118
\mathchardef\pi="7119
\mathchardef\rho="711A
\mathchardef\sigma="711B
\mathchardef\tau="711C
\mathchardef\phi="711E
\mathchardef\chi="711F
\mathchardef\psi="7120
\mathchardef\omega="7121
\mathchardef\varepsilon="7122
\mathchardef\vartheta="7123
\mathchardef\varpi="7124
\mathchardef\varrho="7125
\mathchardef\varsigma="7126
\mathchardef\varphi="7127


\def\beginlinemode{\endmode
  \begingroup\parskip=0pt \obeylines\def\\{\par}\def\endmode{\par\endgroup}}
\def\beginparmode{\endmode
  \begingroup \def\endmode{\par\endgroup}}
\let\endmode=\par
{\obeylines\gdef\
{}}
\def\singlespace{\baselineskip=\normalbaselineskip}

\def\oneandahalfspace{\baselineskip=\normalbaselineskip
  \multiply\baselineskip by 3 \divide\baselineskip by 2}
\def\doublespace{\baselineskip=\normalbaselineskip \multiply\baselineskip by 2}

\newcount\firstpageno
\firstpageno=2
\footline={\ifnum\pageno<\firstpageno{\hfil}\else{\hfil\twelverm\folio\hfil}\fi}
\def\toppageno{\global\footline={\hfil}\global\headline
  ={\ifnum\pageno<\firstpageno{\hfil}\else{\hfil\twelverm\folio\hfil}\fi}}
\let\rawfootnote=\footnote		
\def\footnote#1#2{{\rm\singlespace\parindent=0pt\parskip=0pt
  \rawfootnote{#1}{#2\hfill\vrule height 0pt depth 6pt width 0pt}}}
\def\raggedcenter{\leftskip=4em plus 12em \rightskip=\leftskip
  \parindent=0pt \parfillskip=0pt \spaceskip=.3333em \xspaceskip=.5em
  \pretolerance=9999 \tolerance=9999
  \hyphenpenalty=9999 \exhyphenpenalty=9999 }
\def\dateline{\rightline{\ifcase\month\or
  January\or February\or March\or April\or May\or June\or
  July\or August\or September\or October\or November\or December\fi
  \space\number\year}}
\def\received{\vskip 3pt plus 0.2fill
 \centerline{\sl (Received\space\ifcase\month\or
  January\or February\or March\or April\or May\or June\or
  July\or August\or September\or October\or November\or December\fi
  \qquad, \number\year)}}


\hsize=6.5truein
\hoffset=0pt
\vsize=8.9truein
\voffset=0pt
\parskip=\medskipamount
\def\\{\cr}
\twelvepoint		
\doublespace		
\overfullrule=0pt	




\def\title			
  {\null\vskip 3pt plus 0.2fill
   \beginlinemode \doublespace \raggedcenter \bf}

\def\author			
  {\vskip 3pt plus 0.2fill \beginlinemode
   \singlespace \raggedcenter\sc}

\def\affil			
  {\vskip 3pt plus 0.1fill \beginlinemode
   \oneandahalfspace \raggedcenter \sl}

\def\abstract			
  {\vskip 3pt plus 0.3fill \beginparmode
   \oneandahalfspace ABSTRACT: }

\def\endtitlepage		
  {\endpage			
   \body}
\let\endtopmatter=\endtitlepage

\def\body			
  {\beginparmode}		

\def\head#1{			
  \goodbreak\vskip 0.5truein	
  {\immediate\write16{#1}
   \raggedcenter \uppercase{#1}\par}
   \nobreak\vskip 0.25truein\nobreak}

\def\subhead#1{			
  \vskip 0.25truein		
  {\raggedcenter {#1} \par}
   \nobreak\vskip 0.25truein\nobreak}

\def\beginitems{
\par\medskip\bgroup\def\i##1 {\item{##1}}\def\ii##1 {\itemitem{##1}}
\leftskip=36pt\parskip=0pt}
\def\enditems{\par\egroup}

\def\beneathrel#1\under#2{\mathrel{\mathop{#2}\limits_{#1}}}

\def\refto#1{$^{#1}$}		

\def\references			
  {\head{References}		
   \beginparmode
   \frenchspacing \parindent=0pt \leftskip=1truecm
   \parskip=8pt plus 3pt \everypar{\hangindent=\parindent}}

\gdef\refis#1{\item{#1.\ }}			

\gdef\journal#1, #2, #3, 1#4#5#6{		
    {\sl #1~}{\bf #2}, #3 (1#4#5#6)}		

\def\prd{\journal Phys. Rev. D, }

\def\prl{\journal Phys. Rev. Lett., }

\def\cmp{\journal Comm. Math. Phys., }

\def\endreferences{\body}

\def\figurecaptions		
  {\endpage
   \beginparmode
   \head{Figure Captions}
}

\def\endpage			
  {\vfill\eject}

\def\endpaper			
  {\endmode\vfill\supereject}


\def\heading				
  {\vskip 0.5truein plus 0.1truein	
   \beginparmode \def\\{\par} \parskip=0pt \singlespace \raggedcenter}

\def\subheading				
  {\vskip 0.25truein plus 0.1truein	
   \beginlinemode \singlespace \parskip=0pt \def\\{\par}\raggedcenter}

\def\tag#1$${\eqno(#1)$$}

\def\align#1$${\eqalign{#1}$$}

\def\aligntag#1$${\gdef\tag##1\\{&(##1)\cr}\eqalignno{#1\\}$$
  \gdef\tag##1$${\eqno(##1)$$}}

\def\endaligntag{}

\def\overset #1\to#2{{\mathop{#2}\limits^{#1}}}
\def\underset#1\to#2{{\let\next=#1\mathpalette\undersetpalette#2}}
\def\undersetpalette#1#2{\vtop{\baselineskip0pt
\ialign{$\mathsurround=0pt #1\hfil##\hfil$\crcr#2\crcr\next\crcr}}}


\def\ref#1{Ref.~#1}			
\def\Ref#1{Ref.~#1}			
\def\[#1]{[\cite{#1}]}
\def\cite#1{{#1}}
\let\eq=\Eq\let\eqs=\Eqs		
\def\(#1){(\call{#1})}
\def\call#1{{#1}}
\def\taghead#1{}
\def\frac#1#2{{#1 \over #2}}
\def\half{{\frac 12}}

\def\12{{1\over2}}

\def\ie{{\it i.e.,\ }}

\def\cf{{\sl cf.\ }}
\def\sla{\raise.15ex\hbox{$/$}\kern-.57em}
\def\leaderfill{\leaders\hbox to 1em{\hss.\hss}\hfill}
\def\twiddle{\lower.9ex\rlap{$\kern-.1em\scriptstyle\sim$}}
\def\bigtwiddle{\lower1.ex\rlap{$\sim$}}
\def\gtwid{\mathrel{\raise.3ex\hbox{$>$\kern-.75em\lower1ex\hbox{$\sim$}}}}
\def\ltwid{\mathrel{\raise.3ex\hbox{$<$\kern-.75em\lower1ex\hbox{$\sim$}}}}
\def\square{\kern1pt\vbox{\hrule height 1.2pt\hbox{\vrule width 1.2pt\hskip 3pt
   \vbox{\vskip 6pt}\hskip 3pt\vrule width 0.6pt}\hrule height 0.6pt}\kern1pt}
\def\tdot#1{\mathord{\mathop{#1}\limits^{\kern2pt\ldots}}}

\def\pmb#1{\setbox0=\hbox{#1}%
  \kern-.025em\copy0\kern-\wd0
  \kern  .05em\copy0\kern-\wd0
  \kern-.025em\raise.0433em\box0 }

\catcode`@=11
\newcount\r@fcount \r@fcount=0
\newcount\r@fcurr
\immediate\newwrite\reffile
\newif\ifr@ffile\r@ffilefalse
\def\w@rnwrite#1{\ifr@ffile\immediate\write\reffile{#1}\fi\message{#1}}

\def\writer@f#1>>{}
\def\referencefile{
  \r@ffiletrue\immediate\openout\reffile=\jobname.ref%
  \def\writer@f##1>>{\ifr@ffile\immediate\write\reffile%
    {\noexpand\refis{##1} = \csname r@fnum##1\endcsname = %
     \expandafter\expandafter\expandafter\strip@t\expandafter%
     \meaning\csname r@ftext\csname r@fnum##1\endcsname\endcsname}\fi}%
  \def\strip@t##1>>{}}

\def\citeall#1{\xdef#1##1{#1{\noexpand\cite{##1}}}}
\def\cite#1{\each@rg\citer@nge{#1}}	

\def\each@rg#1#2{{\let\thecsname=#1\expandafter\first@rg#2,\end,}}
\def\first@rg#1,{\thecsname{#1}\apply@rg}	
\def\apply@rg#1,{\ifx\end#1\let\next=\relax
\else,\thecsname{#1}\let\next=\apply@rg\fi\next}

\def\citer@nge#1{\citedor@nge#1-\end-}	
\def\citer@ngeat#1\end-{#1}
\def\citedor@nge#1-#2-{\ifx\end#2\r@featspace#1 
  \else\citel@@p{#1}{#2}\citer@ngeat\fi}	
\def\citel@@p#1#2{\ifnum#1>#2{\errmessage{Reference range #1-#2\space is bad.}%
    \errhelp{If you cite a series of references by the notation M-N, then M and
    N must be integers, and N must be greater than or equal to M.}}\else%
 {\count0=#1\count1=#2\advance\count1 by1\relax\expandafter\r@fcite\the\count0,%
  \loop\advance\count0 by1\relax
    \ifnum\count0<\count1,\expandafter\r@fcite\the\count0,%
  \repeat}\fi}

\def\r@featspace#1#2 {\r@fcite#1#2,}	
\def\r@fcite#1,{\ifuncit@d{#1}
    \newr@f{#1}%
    \expandafter\gdef\csname r@ftext\number\r@fcount\endcsname%
                     {\message{Reference #1 to be supplied.}%
                      \writer@f#1>>#1 to be supplied.\par}%
 \fi%
 \csname r@fnum#1\endcsname}
\def\ifuncit@d#1{\expandafter\ifx\csname r@fnum#1\endcsname\relax}%
\def\newr@f#1{\global\advance\r@fcount by1%
    \expandafter\xdef\csname r@fnum#1\endcsname{\number\r@fcount}}

\let\r@fis=\refis			
\def\refis#1#2#3\par{\ifuncit@d{#1}
   \newr@f{#1}%
   \w@rnwrite{Reference #1=\number\r@fcount\space is not cited up to now.}\fi%
  \expandafter\gdef\csname r@ftext\csname r@fnum#1\endcsname\endcsname%
  {\writer@f#1>>#2#3\par}}

\def\ignoreuncited{
   \def\refis##1##2##3\par{\ifuncit@d{##1}%
     \else\expandafter\gdef\csname r@ftext\csname r@fnum##1\endcsname\endcsname%
     {\writer@f##1>>##2##3\par}\fi}}

\def\r@ferr{\endreferences\errmessage{I was expecting to see
\noexpand\endreferences before now;  I have inserted it here.}}
\let\r@ferences=\references
\def\references{\r@ferences\def\endmode{\r@ferr\par\endgroup}}

\let\endr@ferences=\endreferences
\def\endreferences{\r@fcurr=0
  {\loop\ifnum\r@fcurr<\r@fcount
    \advance\r@fcurr by 1\relax\expandafter\r@fis\expandafter{\number\r@fcurr}%
    \csname r@ftext\number\r@fcurr\endcsname%
  \repeat}\gdef\r@ferr{}\endr@ferences}


\let\r@fend=\endpaper\gdef\endpaper{\ifr@ffile
\immediate\write16{Cross References written on []\jobname.REF.}\fi\r@fend}

\catcode`@=12

\citeall\refto		
\citeall\ref		%
\citeall\Ref		%

\catcode`@=11
\newcount\tagnumber\tagnumber=0

\immediate\newwrite\eqnfile
\newif\if@qnfile\@qnfilefalse
\def\write@qn#1{}
\def\writenew@qn#1{}
\def\w@rnwrite#1{\write@qn{#1}\message{#1}}
\def\@rrwrite#1{\write@qn{#1}\errmessage{#1}}

\def\taghead#1{\gdef\t@ghead{#1}\global\tagnumber=0}
\def\t@ghead{}

\expandafter\def\csname @qnnum-3\endcsname
  {{\t@ghead\advance\tagnumber by -3\relax\number\tagnumber}}
\expandafter\def\csname @qnnum-2\endcsname
  {{\t@ghead\advance\tagnumber by -2\relax\number\tagnumber}}
\expandafter\def\csname @qnnum-1\endcsname
  {{\t@ghead\advance\tagnumber by -1\relax\number\tagnumber}}
\expandafter\def\csname @qnnum0\endcsname
  {\t@ghead\number\tagnumber}
\expandafter\def\csname @qnnum+1\endcsname
  {{\t@ghead\advance\tagnumber by 1\relax\number\tagnumber}}
\expandafter\def\csname @qnnum+2\endcsname
  {{\t@ghead\advance\tagnumber by 2\relax\number\tagnumber}}
\expandafter\def\csname @qnnum+3\endcsname
  {{\t@ghead\advance\tagnumber by 3\relax\number\tagnumber}}

\def\equationfile{%
  \@qnfiletrue\immediate\openout\eqnfile=\jobname.eqn%
  \def\write@qn##1{\if@qnfile\immediate\write\eqnfile{##1}\fi}
  \def\writenew@qn##1{\if@qnfile\immediate\write\eqnfile
    {\noexpand\tag{##1} = (\t@ghead\number\tagnumber)}\fi}
}

\def\callall#1{\xdef#1##1{#1{\noexpand\call{##1}}}}
\def\call#1{\each@rg\callr@nge{#1}}

\def\each@rg#1#2{{\let\thecsname=#1\expandafter\first@rg#2,\end,}}
\def\first@rg#1,{\thecsname{#1}\apply@rg}
\def\apply@rg#1,{\ifx\end#1\let\next=\relax%
\else,\thecsname{#1}\let\next=\apply@rg\fi\next}

\def\callr@nge#1{\calldor@nge#1-\end-}
\def\callr@ngeat#1\end-{#1}
\def\calldor@nge#1-#2-{\ifx\end#2\@qneatspace#1 %
  \else\calll@@p{#1}{#2}\callr@ngeat\fi}
\def\calll@@p#1#2{\ifnum#1>#2{\@rrwrite{Equation range #1-#2\space is bad.}
\errhelp{If you call a series of equations by the notation M-N, then M and
N must be integers, and N must be greater than or equal to M.}}\else%
 {\count0=#1\count1=#2\advance\count1 by1\relax\expandafter\@qncall\the\count0,%
  \loop\advance\count0 by1\relax%
    \ifnum\count0<\count1,\expandafter\@qncall\the\count0,%
  \repeat}\fi}

\def\@qneatspace#1#2 {\@qncall#1#2,}
\def\@qncall#1,{\ifunc@lled{#1}{\def\next{#1}\ifx\next\empty\else
  \w@rnwrite{Equation number \noexpand\(>>#1<<) has not been defined yet.}
  >>#1<<\fi}\else\csname @qnnum#1\endcsname\fi}

\let\eqnono=\eqno
\def\eqno(#1){\tag#1}
\def\tag#1$${\eqnono(\displayt@g#1 )$$}

\def\aligntag#1\endaligntag
  $${\gdef\tag##1\\{&(##1 )\cr}\eqalignno{#1\\}$$
  \gdef\tag##1$${\eqnono(\displayt@g##1 )$$}}

\def\eqalignno#1{\displ@y \tabskip\centering
  \halign to\displaywidth{\hfil$\displaystyle{##}$\tabskip\z@skip
    &$\displaystyle{{}##}$\hfil\tabskip\centering
    &\llap{$\displayt@gpar##$}\tabskip\z@skip\crcr
    #1\crcr}}

\def\displayt@gpar(#1){(\displayt@g#1 )}

\def\displayt@g#1 {\rm\ifunc@lled{#1}\global\advance\tagnumber by1
        {\def\next{#1}\ifx\next\empty\else\expandafter
        \xdef\csname @qnnum#1\endcsname{\t@ghead\number\tagnumber}\fi}%
  \writenew@qn{#1}\t@ghead\number\tagnumber\else
        {\edef\next{\t@ghead\number\tagnumber}%
        \expandafter\ifx\csname @qnnum#1\endcsname\next\else
        \w@rnwrite{Equation \noexpand\tag{#1} is a duplicate number.}\fi}%
  \csname @qnnum#1\endcsname\fi}

\def\ifunc@lled#1{\expandafter\ifx\csname @qnnum#1\endcsname\relax}

\let\@qnend=\end\gdef\end{\if@qnfile
\immediate\write16{Equation numbers written on []\jobname.EQN.}\fi\@qnend}

\catcode`@=12


\def\3he{{$^3${\rm He}}}

\def\ie{{\it i.e.,\ }}

\def\slD{\raise.15ex\hbox{$/$}\kern-.53em\hbox{$D$}}
\def\dsl{\raise.15ex\hbox{$/$}\kern-.57em\hbox{$\Delta$}}
\def\slp{{\raise.15ex\hbox{$/$}\kern-.57em\hbox{$\partial$}}}
\def\nsl{\raise.15ex\hbox{$/$}\kern-.57em\hbox{$\nabla$}}
\def\sla{\raise.15ex\hbox{$/$}\kern-.57em\hbox{$\rightarrow$}}
\def\slla{\raise.15ex\hbox{$/$}\kern-.57em\hbox{$\lambda$}}
\def\slb{\raise.15ex\hbox{$/$}\kern-.57em\hbox{$b$}}
\def\lnp{\raise.15ex\hbox{$/$}\kern-.57em\hbox{$p$}}
\def\lnk{\raise.15ex\hbox{$/$}\kern-.57em\hbox{$k$}}
\def\lnK{\raise.15ex\hbox{$/$}\kern-.57em\hbox{$K$}}
\def\lnq{\raise.15ex\hbox{$/$}\kern-.57em\hbox{$q$}}


\def\pmb#1{\setbox0=\hbox{$#1$}%
\kern-.025em\copy0\kern-\wd0
\kern.05em\copy0\kern-\wd0
\kern-.025em\raise.0433em\box0 }

\def\q2{{Q^2}}
\def\gtwid{\raise.3ex\hbox{$>$\kern-.75em\lower1ex\hbox{$\sim$}}}
\def\ltwid{\raise.3ex\hbox{$<$\kern-.75em\lower1ex\hbox{$\sim$}}}
\def\12{{1\over2}}
\def\part{\partial}

\def\low#1{\lower.5ex\hbox{${}_#1$}}

\def\psl{\raise.15ex\hbox{$/$}\kern-.57em\hbox{$\partial$}}
\def\partt{\raise.15ex\hbox{$\widetilde$}{\kern-.37em\hbox{$\partial$}}}

\def\topppageno1{\global\footline={\hfil}\global\headline
={\ifnum\pageno<\firstpageno{\hfil}\else{\hss\twelverm --\ \folio
\ --\hss}\fi}}
 
\def\toppageno2{\global\footline={\hfil}\global\headline
={\ifnum\pageno<\firstpageno{\hfil}\else{\rightline{\hfill\hfill
\twelverm \ \folio
\ \hss}}\fi}}

\def\prl#1{Phys.\ Rev.\ Lett.\ {\bf #1}}
\def\prd#1{Phys.\ Rev.\ {\bf D#1}}
\def\plb#1{Phys.\ Lett.\ {\bf B#1}}
\def\npb#1{Nucl.\ Phys.\ {\bf B#1}}
\def\cmp#1{Comm.\ Math.\ Phys.\ {\bf #1}}

\def\ie{{\it i.e.},\ }

\def\ref#1{${}^{#1}$}
\def\nsection#1 #2{\leftline{\rlap{#1}\indent\relax #2}}

\def\prl#1{Phys.\ Rev.\ Lett.\ {\bf #1}}
\def\prd#1{Phys.\ Rev.\ {\bf D#1}}
\def\plb#1{Phys.\ Lett.\ {\bf #1B}}
\def\npb#1{Nucl.\ Phys.\ {\bf B#1}}

\newread\epsffilein    
\newif\ifepsffileok    
\newif\ifepsfbbfound   
\newif\ifepsfverbose   
\newif\ifepsfdraft     
\newdimen\epsfxsize    
\newdimen\epsfysize    
\newdimen\epsftsize    
\newdimen\epsfrsize    
\newdimen\epsftmp      
\newdimen\pspoints     
\pspoints=1bp          
\epsfxsize=0pt         
\epsfysize=0pt         
\def\epsfbox#1{\global\def\epsfllx{72}\global\def\epsflly{72}%
   \global\def\epsfurx{540}\global\def\epsfury{720}%
   \def\lbracket{[}\def\testit{#1}\ifx\testit\lbracket
   \let\next=\epsfgetlitbb\else\let\next=\epsfnormal\fi\next{#1}}%
\def\epsfgetlitbb#1#2 #3 #4 #5]#6{\epsfgrab #2 #3 #4 #5 .\\%
   \epsfsetgraph{#6}}%
\def\epsfnormal#1{\epsfgetbb{#1}\epsfsetgraph{#1}}%
\def\epsfgetbb#1{%
%
%
\openin\epsffilein=#1
\ifeof\epsffilein\errmessage{I couldn't open #1, will ignore it}\else
%
%
   {\epsffileoktrue \chardef\other=12
    \def\do##1{\catcode`##1=\other}\dospecials \catcode`\ =10
    \loop
       \read\epsffilein to \epsffileline
       \ifeof\epsffilein\epsffileokfalse\else
%
%
          \expandafter\epsfaux\epsffileline:. \\%
       \fi
   \ifepsffileok\repeat
   \ifepsfbbfound\else
    \ifepsfverbose\message{No bounding box comment in #1; using defaults}\fi\fi
   }\closein\epsffilein\fi}%
%
%
%
\def\epsfclipoff{\def\epsfclipstring{\ifepsfdraft\space clip\fi}}%
\epsfclipoff
\def\epsfsetgraph#1{%
   \epsfrsize=\epsfury\pspoints
   \advance\epsfrsize by-\epsflly\pspoints
   \epsftsize=\epsfurx\pspoints
   \advance\epsftsize by-\epsfllx\pspoints
%
%
   \epsfxsize\epsfsize\epsftsize\epsfrsize
   \ifnum\epsfxsize=0 \ifnum\epsfysize=0
      \epsfxsize=\epsftsize \epsfysize=\epsfrsize
      \epsfrsize=0pt
%
%
     \else\epsftmp=\epsftsize \divide\epsftmp\epsfrsize
       \epsfxsize=\epsfysize \multiply\epsfxsize\epsftmp
       \multiply\epsftmp\epsfrsize \advance\epsftsize-\epsftmp
       \epsftmp=\epsfysize
       \loop \advance\epsftsize\epsftsize \divide\epsftmp 2
       \ifnum\epsftmp>0
          \ifnum\epsftsize<\epsfrsize\else
             \advance\epsftsize-\epsfrsize \advance\epsfxsize\epsftmp \fi
       \repeat
       \epsfrsize=0pt
     \fi
   \else \ifnum\epsfysize=0
     \epsftmp=\epsfrsize \divide\epsftmp\epsftsize
     \epsfysize=\epsfxsize \multiply\epsfysize\epsftmp   
     \multiply\epsftmp\epsftsize \advance\epsfrsize-\epsftmp
     \epsftmp=\epsfxsize
     \loop \advance\epsfrsize\epsfrsize \divide\epsftmp 2
     \ifnum\epsftmp>0
        \ifnum\epsfrsize<\epsftsize\else
           \advance\epsfrsize-\epsftsize \advance\epsfysize\epsftmp \fi
     \repeat
     \epsfrsize=0pt
    \else
     \epsfrsize=\epsfysize
    \fi
   \fi
%
%
   \ifepsfverbose\message{#1: width=\the\epsfxsize, height=\the\epsfysize}\fi
   \epsftmp=10\epsfxsize \divide\epsftmp\pspoints
   \vbox to\epsfysize{\vfil\hbox to\epsfxsize{%
      \ifnum\epsfrsize=0\relax
        \includegraphics{\ifepsfdraft}%
      \else
        \epsfrsize=10\epsfysize \divide\epsfrsize\pspoints
        \includegraphics{\ifepsfdraft}%
      \fi
      \hfil}}%
\global\epsfxsize=0pt\global\epsfysize=0pt}%
%
%
{\catcode`\%=12 \global\let\epsfpercent=
%
%
\long\def\epsfaux#1#2:#3\\{\ifx#1\epsfpercent
   \def\testit{#2}\ifx\testit\epsfbblit
      \epsfgrab #3 . . . \\%
      \epsffileokfalse
      \global\epsfbbfoundtrue
   \fi\else\ifx#1\par\else\epsffileokfalse\fi\fi}%
%
%
\def\epsfempty{}%
\def\epsfgrab #1 #2 #3 #4 #5\\{%
\global\def\epsfllx{#1}\ifx\epsfllx\epsfempty
      \epsfgrab #2 #3 #4 #5 .\\\else
   \global\def\epsflly{#2}%
   \global\def\epsfurx{#3}\global\def\epsfury{#4}\fi}%
%
%
\def\epsfsize#1#2{\epsfxsize}
%
%

\def\exp{{\rm exp\;}}
\def\tr{{\rm tr\;}}
\def\square#1#2{{\vcenter{\vbox{\hrule height.#2pt
    \hbox{\vrule width.#2pt height#1pt \kern#1pt \vrule width.#2pt}
    \hrule height.#2pt}}}}
\def\Box{{\mathchoice{\sl\square63}{\sl\square63}\square{2.1}3\square{1.5}3}\,}
\def\psibar{{\overline{\psi}}}
\def\tk{{\tilde{\kappa}}}
\def\phat{{\hat{p}}}
\def\e{{\rm e}}
\def\tlp{{\tilde{p}}}
\def\tlk{{\tilde{k}}}
\def\slash{{/\!\!\!}}

{\parindent=0pt December 1997 \hfill{HU-EP-97/95, 
TAUP-2670-97, Wash. U. HEP/97-61} }

\title Chiral Fermions on the Lattice through Gauge Fixing -- Perturbation 
Theory

\author Wolfgang Bock$^a$\footnote{${}^*$}{e-mail: bock@%
linde.physik.hu-berlin.de}, %
Maarten F.L. Golterman$^b$\footnote{${}^{**}$}{e-mail: maarten@%
aapje.wustl.edu}
and Yigal Shamir$^c$\footnote{${}^\dagger$}%
{e-mail: ftshamir@wicc.weizmann.ac.il}
\affil
$^a$Institute of Physics, Humboldt University 
Invalidenstr.~110,~10099~Berlin,~Germany
\null
$^b$Department of Physics, Washington University
St.~Louis,~MO~63130,~USA
\null
$^c$School of Physics and Astronomy, Tel-Aviv University 
Ramat~Aviv,~69978~Israel

\abstract{
We study the gauge-fixing approach to the construction of lattice chiral
gauge theories in one-loop weak-coupling perturbation theory.  We show how
infrared properties of the gauge degrees of freedom determine the nature of
the continuous phase transition at which we take the continuum limit.
The fermion self-energy and the vacuum polarization are calculated, and
confirm that, in the abelian case, this approach can be used to put
chiral gauge theories on the lattice in four dimensions.  We comment on the 
generalization to the nonabelian case. }
\endtopmatter

\subhead{\bf 1. Introduction}

In a recent paper [\cite{bgslett}] we have shown that one can construct
models with chiral fermions on the lattice by using a lattice action which
contains a discretization of a covariant continuum gauge-fixing term.
The model we investigated is a concrete implementation [\cite{golsha}]
of the so-called ``Rome approach" [\cite{rome,sha}].  

In lattice chiral gauge theories the gauge symmetry is explicitly broken
for nonzero values of the lattice spacing, even in anomaly-free models.
The basic reason for this is that each fermion species has to contribute
its part to the chiral anomaly, and in order to do so, chiral symmetry
has to be explicitly broken in the regulated theory [\cite{karsmit}]
(see also ref. [\cite{shalat95}] and references therein).  On the lattice,
the gauge-symmetry breaking induced by quantum effects
is not restricted to the anomaly, but includes
infinitely many higher-dimensional operators which are suppressed 
by powers of the lattice spacing (are ``irrelevant") for smooth external
gauge fields.  However, for arbitrarily ``rough" lattice
gauge fields, these operators potentially lead to unsuppressed interactions
between the fermions and the gauge degrees of freedom (the longitudinal
modes of the gauge field).  Typically, this phenomenon alters the fermion
spectrum of the theory nonperturbatively, leading to a vectorlike rather
than a chiral fermion content in the continuum limit (for reviews, see
refs. [\cite{petcher,shalat95}]).

In order to remedy this problem, it is natural to consider gauge-fixed
lattice gauge theories [\cite{rome,sha}].  
It was argued in ref. [\cite{sha}] that a smooth
gauge may lead to a suppression of rough lattice gauge
fields such that a location in the phase diagram of the theory exists
where the fermion spectrum remains chiral.  In
this case, both the transversal and longitudinal modes are controlled by the
bare lattice gauge coupling, so that the lattice theory can be
systematically studied in weak-coupling perturbation theory.

In order for the lattice theory to admit a perturbative expansion, the
gauge-fixing action should have a global minimum at the perturbative
vacuum, $A_\mu=0$.  A discretization of the standard Lorentz gauge-fixing
term with this property was proposed in ref. [\cite{golsha}].   
A simplified version of this model was then studied nonperturbatively for the
abelian case.  In this ``reduced" model only the longitudinal modes of the
gauge field (or, equivalently, the gauge degrees of freedom) are taken into
account.  Since these are precisely the degrees of freedom that, without gauge
fixing, destroy the chiral nature of the fermions, it is important to study
such reduced models first, in order to demonstrate that the fermions remain
chiral despite their interactions with the gauge degrees of freedom.

In refs. [\cite{sha,golsha}] it was argued that, for small gauge coupling,
the gauge-fixed lattice action leads to a continuous phase transition between a
Higgs phase, and a novel ``directional" phase, in which the gauge
field condenses.  At the phase transition (which belongs to a universality
class different
from the usual Higgs transition), the gauge field is massless, and
a continuum limit can be defined.  The existence of this phase transition was
confirmed in the reduced abelian model by high-statistics numerical computations
and in the mean-field approximation [\cite{bgs}]. In the reduced model,
which is always invariant under constant gauge transformations, the Higgs phase
corresponds to a phase with broken symmetry, which however gets restored
at the phase transition between the Higgs and ``directional" phases.  
This symmetry restoration  is of crucial importance, since
it allows us to unambiguously determine the fermionic quantum numbers
under the (global remnant of the) gauge group.
Using Wilson fermions, the existence of 
undoubled fermions in the desired chiral representation of the gauge
group was confirmed numerically in ref.  [\cite{bgslett}].

In this paper, we study the reduced model in detail in weak-coupling 
perturbation
theory.  In section 2, we define the fully gauged and reduced models, and
explain how perturbation theory may be set up systematically.  In section 3,
where we limit ourselves to the abelian case, we show how the dynamics of
the gauge degrees of freedom leads to the continuous phase transition
mentioned above, and how the symmetry gets restored at the phase
transition. In section 4, we discuss the one-loop fermion self-energy, 
and demonstrate that indeed free chiral fermions with the
correct quantum numbers emerge at this phase transition in the reduced model.
We then go on to discuss the vacuum polarization in section 5.  
We calculate the shift
in the location of the phase transition induced by the fermions at one
loop. We show that, at the phase transition, the gauge degrees of
freedom decouple from the fermions (a result that also follows from, and
is consistent with, the
fermion self-energy calculated in section 4), and that the
expected fermionic contribution to the
$\beta$-function is obtained for the gauge coupling.  All these
results confirm that, at least in the abelian case, our lattice theory
leads to the desired chiral gauge theory when the continuum limit is
taken at the continuous phase transition at weak gauge coupling. Some
of the results of this paper have already been used in a comparison with
the numerical results of refs. [\cite{bgslett,bgs}]. In
section 6, we discuss the issue of fermion number nonconservation at the
level of perturbation theory. Following ref. [\cite{dugman}], we show that 
a gauge invariant fermion-number current can be constructed with the
correct anomaly in the continuum limit.  In the last section, we summarize
our results, and outline some of the most important open problems. 
We refer to refs. [\cite{lat97i,lat97ii}] for a less technical account of
our work.

\subhead{\bf 2. The model}

Let us start with the action for the fully gauged lattice chiral fermion
theory.  We will assume that all physical fermions are left-handed, and that 
they transform in some (not necessarily irreducible) representation of a gauge 
group $G$.  This representation will have to be anomaly-free if
a unitary continuum limit is to exist.  The complete action can be
written as a sum of terms, each of which we will introduce below:
$$S_V=S_{\rm plaq}+S_{\rm gf}+S_{\rm ghost}+S_{\rm fermion}+S_{\rm ct}.
\eqno(SV)$$
For $S_{\rm plaq}$ we will assume the usual plaquette term with the link
variables $U_{x,\mu}=\exp(iA_{x,\mu})$ in the fundamental representation.  
For $S_{\rm gf}$ we will take the lattice version of the square of the
Lorentz gauge condition that we proposed in ref. [\cite{golsha}]:
$$S_{\rm gf}={1\over{2\xi g^2}}\;\tr
\left(\sum_{xyz}\Box_{xy}(U)\;\Box_{yz}(U)-\sum_x
B_x^2(V(U))\right),\eqno(Sgf)$$
where
$$\Box_{xy}(U)=\sum_\mu(\delta_{x+\mu,y}U_{x,\mu}+\delta_{x-\mu,y}
U^\dagger_{y,\mu})-8\delta_{x,y}\eqno(covlapl)$$
is the covariant lattice laplacian, and 
$$B_x(V)=\sum_\mu\left({{V_{x-\mu,\mu}+V_{x,\mu}}\over 2}\right)^2,\eqno(B)$$
with 
$$V_{x,\mu}={1\over{2i}}(U_{x,\mu}-U^\dagger_{x,\mu})=A_{x,\mu}+O(A^3).
\eqno(V)$$
$g$ is the bare gauge coupling, and $\xi$ is the bare gauge-fixing
parameter.  It is straightforward to show that, in the classical
continuum limit,
$$S_{\rm gf}={1\over{2\xi g^2}}\tr(\partial_\mu
A_{x,\mu})^2+{\rm irrelevant\ operators}.\eqno(Sgfcontl)$$
Of course there are many possible choices for $S_{\rm gf}$ with the same
classical continuum limit.  Our choice here is motivated by two
important properties obeyed by \eq{Sgf}\ [\cite{golsha}]:

\item{$\bullet$}
$S_{\rm gf}$ has a unique absolute minimum at $U_{x,\mu}=I$, validating
weak-coupling perturbation theory in $g$.
\item{$\bullet$}
Our choice of $S_{\rm gf}$ leads to a critical behavior suitable for 
taking a continuum limit in the limit $g\to 0$.

Both properties will be used and discussed in this paper. The fact that
this gauge-fixing action has a unique minimum is closely related to the
fact that, on the lattice, 
it is not the square of a local gauge-fixing condition.  As a result,
the action $S_V$ (even without the fermions) is not BRST invariant.  This
situation allows us to avoid a theorem stating that expectation values
of gauge-invariant operators would vanish in a lattice model with exact
BRST invariance, due to the existence of lattice Gribov copies in such
lattice models [\cite{neu}]. 

In the BRST approach, the gauge-fixing part of the action is not complete 
without a Fadeev--Popov term $S_{\rm ghost}$. However, we will not specify 
this term here, as we
will be mostly concerned with the abelian case $G=U(1)$, or with
one-loop calculations not involving ghost loops.

For the fermion action, we will choose to use Wilson fermions.  For each
left-handed fermion $\psi_L$ we introduce a right-handed ``spectator"
fermion $\psi_R$.  This allows us to construct a Wilson term that
will serve to remove the fermion doublers, of course at the expense of
gauge invariance [\cite{smiswi}].  The fermion action is
$$\eqalignno{
S_{\rm fermion}=\half
\sum_{x,\mu}\Bigl(&\psibar_{x}\gamma_\mu(U_{x,\mu}P_L+P_R)\psi_{x+\mu}
-\psibar_{x+\mu}\gamma_\mu(U^\dagger_{x,\mu}P_L+P_R)\psi_{x}\cr
&-r(\psibar_x\psi_{x+\mu}+\psibar_{x+\mu}\psi_x-2\psibar_x\psi_x)\Bigr).
&(Sfermion)}$$
$P_{L(R)}$ are the left(right)-handed projectors $\half(1\mp\gamma_5)$,
and $r$ is the Wilson parameter.  Since the Wilson term breaks the
left-handed $G$-invariance anyway, we choose to not put any gauge fields
in, and $S_{\rm fermion}$ is therefore invariant under the shift symmetry
[\cite{golpet}]
$$\psi_R\to\psi_R+\epsilon_R,\ \ \ \psibar_R\to\psibar_R+
\overline{\epsilon}_R\;.\eqno(shiftsymm)$$

Since gauge invariance (or more precisely, BRST invariance)
is broken by the fermion action and by the 
gauge-fixing action, we will need to add counterterms, $S_{\rm ct}$.
In principle, all relevant and marginal counterterms which are allowed
by the exact symmetries of the lattice theory will be needed 
[\cite{rome}].  The most important one for our purposes in this paper is
the gauge-boson mass counterterm, which is the only dimension-two
counterterm.  All other counterterms are of dimension four, since a
fermion-mass counterterm is forbidden by shift symmetry (lower
dimension counterterms involving ghost fields are excluded by lattice 
symmetries as well [\cite{rome}]).  So we will choose 
$$S_{\rm ct}=-\kappa\;\tr\sum_{x,\mu}\left(U_{x,\mu}+U^\dagger_{x,\mu}\right)
+{\rm marginal\ terms},\eqno(Sct)$$
where we do not need to specify the marginal terms for this paper.  They
could be constructed from their continuum form, by replacing
$A_{x,\mu}\to V_{x,\mu}$ (\cf\ \eq{V}) and partial derivatives by
difference operators.

Since the action $S_V$ is not gauge invariant, we may introduce a 
St\"uckelberg field $\phi_x\in G$, and write the action as 
$$S_H=S_{\rm plaq}+S^\phi_{\rm gf}+S^\phi_{\rm ghost}+S^\phi_{\rm fermion}
+S^\phi_{\rm ct},\eqno(SH)$$
with
$$\eqalignno{
S^\phi_{\rm gf}&={1\over{2\xi g^2}}\;\tr
\sum_x\left(\phi^\dagger_x(\Box^2(U)\phi)_x-B_x^2(V^\phi(U))\right),\cr
S^\phi_{\rm fermion}&=\half
\sum_{x,\mu}\Bigl(\psibar_{x}\gamma_\mu(U_{x,\mu}P_L+P_R)\psi_{x+\mu}
-\psibar_{x+\mu}\gamma_\mu(U^\dagger_{x,\mu}P_L+P_R)\psi_{x}\cr
&\phantom{=}-r((\psibar_x(\phi^\dagger_{x+\mu}P_L+\phi_x P_R)\psi_{x+\mu
}+{\rm h.c.})-2\psibar_x(\phi^\dagger_x P_L+\phi_x P_R)\psi_x)\Bigr),\cr
S^\phi_{\rm ct}&=-\kappa\;\tr\sum_x\phi^\dagger_x(\Box(U)\phi)_x
+{\rm marginal\ terms},&(Sphis)
}$$ 
and in which $V_{x,\mu}$ is replaced by $V^\phi_{x,\mu}$ with
$$V^\phi_{x,\mu}={1\over{2i}}(\phi^\dagger_x U_{x,\mu}\phi_{x+\mu}
-\phi^\dagger_{x+\mu}U^\dagger_{x,\mu}\phi_x).
\eqno(Vphi)$$
Note that $S_{\rm plaq}$ and the $r=0$ part of $S_{\rm fermion}$ do not
change because they are gauge invariant.

$S_H$ is gauge invariant under the transformation
$$\eqalignno{
U_{x,\mu}&\to h_{Lx} U_{x,\mu}h^\dagger_{Lx+\mu},\cr
\phi_x&\to h_{Lx}\phi_x,\cr
\psi_x&\to(h_{Lx} P_L+P_R)\psi_x,&(gaugesym)
}$$
where $h_{Lx}\in G$.  Because of this, $\phi_x$ may be completely
eliminated from $S_H$ by a gauge transformation, and doing so we recover,
as expected,
$$S_V(U,\psi)=S_H(\phi,U,\psi)\Big|_{\phi=I}.\eqno(VHrel)$$
We will refer to $S_{V(H)}$ as the action in the ``vector" (``Higgs")
picture.  The two formulations are entirely equivalent: observables in
the vector picture are mapped into (gauge-invariant) observables in the
Higgs picture, and {\it vice versa} [\cite{foenie}].

Next, we introduce the ``reduced" model, which is obtained from $S_H$ by
setting the {\it gauge field} $U_{x,\mu}$ equal to one.  The reason that
this reduced model is of interest is that, if the full model is to yield
a theory of fermions chirally coupled to gluons in the continuum limit, the 
reduced model should lead to a theory of free chiral fermions (in the
correct representation of the gauge group $G$) in the corresponding
continuum limit.  Ignoring the marginal
counterterms, we obtain the reduced model action
$$\eqalignno{
S_{\rm reduced}=&
\;\tk\;\tr\;\sum_x\left(\phi^\dagger_x(\Box^2\phi)_x-B_x^2(V^{\rm r}(\phi))
\right) -\kappa\;\tr\sum_x\phi^\dagger_x(\Box\phi)_x\cr
&+\half
\sum_{x,\mu}\Bigl(\psibar_{x}\gamma_\mu\psi_{x+\mu}
-\psibar_{x+\mu}\gamma_\mu\psi_{x}&(Sreduced)\cr
&\phantom{=}-r((\psibar_x(\phi^\dagger_{x+\mu}P_L+\phi_x P_R)\psi_{x+\mu
}+{\rm h.c.})-2\psibar_x(\phi^\dagger_x P_L+\phi_x P_R)\psi_x)\Bigr),
}$$
where now $\Box$ is the standard lattice laplacian (\cf\ \eq{covlapl}
with $U_{x,\mu}=I$), 
$$V^{\rm r}_{x,\mu}={1\over{2i}}(\phi^\dagger_x\phi_{x+\mu}
-\phi^\dagger_{x+\mu}\phi_x),\eqno(Vred)$$
and we abbreviated
$$\tk\equiv{1\over{2\xi g^2}}.\eqno(tk)$$
$S_{\rm reduced}$ is invariant under the transformation \eq{gaugesym} for
constant $h_{L x}=h_L$, as well as under the transformation
$$\eqalignno{
\phi&\to\phi h^\dagger_R,\cr
\psi&\to(P_L+h_R P_R)\psi,&(globalsym)
}$$
with $h_R\in G$, \ie\ $S_{\rm reduced}$ has a global $G_L\times G_R$ 
symmetry.  Weak-coupling perturbation theory in $g$ corresponds to
perturbation theory in $1/\tk$.  Note that in the original action in the
vector picture, the gauge-fixing term corresponds to a kinetic term
for the longitudinal part of the gauge field
$U_{x,\mu}$. Therefore $S_V$ is manifestly renormalizable,
and can be treated systematically in perturbation theory in $g$, even though it
is not gauge invariant [\cite{rome,sha}].
In the reduced model, we expand
$$\phi_x=\exp(i\theta_x/\sqrt{2\tk})=\exp(ig\sqrt{\xi}\theta_x),
\eqno(theta)$$
in order to develop perturbation theory.  This
leads to tree-level scalar and fermion propagators $\langle\theta(p)
\theta(q)\rangle=\delta(p+q)G(p)$ and $\langle\psi(p)
\psibar(q)\rangle=\delta(p+q)S(p)$ with
$$\eqalignno{
G(p)=&{1\over{\phat^2(\phat^2+m^2)}},\ \ \ \ \ m^2\equiv{\kappa\over\tk},\cr
S(p)=&\left(i\slash s(p)+rM(p)\right)^{-1}\cr
=&(-i\slash s(p)+rM(p))/D(p),\cr
D(p)=&s^2(p)+r^2M^2(p),&(props)
}$$
where $\phat_\mu=2\sin{(p_\mu/2)}$, $\slash s(p)=\sum_\mu\gamma_\mu
\sin{p_\mu}$, $s^2(p)=\sum_\mu\sin^2{p_\mu}$ and $M(p)=\half\phat^2$.
The vertices can also be read off from $S_{\rm reduced}$ after
expanding $\phi$ in terms of $\theta$.  A vertex with $n$ $\theta$-lines 
has a coupling constant of order $\tk^{-(n-2)/2}$, while a
vertex involving the fermions and $n$ $\theta$-lines has a coupling of
order $\tk^{-n/2}$.  

\subhead{\bf 3. The FM--FMD transition and the continuum limit}

In this section, we will discuss in detail the properties of the phase 
transition that occurs for a critical value $\kappa_c$
of the parameter $\kappa$.  We will assume that $\tk$ is large and
positive (for details on the complete phase diagram, 
see refs. [\cite{bgs,sha,golsha}]). We will limit
ourselves to the case without fermions, and postpone their inclusion
to a later section. We will also simplify the discussion by restricting
ourselves to the abelian case, $G=U(1)$. 

An indication that a continuous phase transition occurs can be obtained
from the $\theta$ propagator (\eq{props}):  if $\kappa<0$, $m^2$ becomes
negative, signaling an instability at $\kappa_c=0$ against the 
condensation of plane waves with nonzero momentum,
which breaks lattice space-time symmetries.  (This value for
$\kappa_c$ is just its tree-level value; its true value will be shifted
by quantum corrections.)  We first observe that, for large $\tk$,
$\phi$ acquires an expectation value, and in fact
$|\langle\phi_x\rangle|\to 1$ for $\tk\to\infty$ (as long as we stay
away from the phase transition line, see below). This 
breaks the global $G_L\times G_R$ symmetry down to the diagonal symmetry
$h_L=h_R$ (\cf\ \eqs{gaugesym,globalsym}).  In order to analyze the situation
for small $|\kappa|$, we substitute
$\phi_x=\exp(iqx)$ into the bosonic part of $S_{\rm reduced}$, which
gives us a potential density $V(q)$:
$$V(q)=\tk\left[4\left(\sum_\mu(1-\cos{q_\mu})\right)^2-
\left(\sum_\mu\sin^2{q_\mu}\right)^2
+2m^2\sum_\mu(1-\cos{q_\mu})\right].\eqno(Vq)$$
It is easy to see that for $m^2>0$, $V(q)\ge 0$ and that
$V(q)=0\Leftrightarrow q=0$. But for $m^2$ negative, the absolute
minimum of $V(q)$ occurs at a nonzero value of $q$; for $m^2$ small
and negative it occurs at [\cite{golsha}]
$$q_\mu=\pm\left({{|m^2|}\over 6}\right)^{1/4},\ \ \ \ \ {\rm all}\ \mu.
\eqno(qsol)$$ 
Hence, for large values of $\tk$, a continuous
phase transition takes place from a
phase with broken symmetry and $q=0$, which we will call the FM 
(ferromagnetic) phase, to a phase with broken symmetry and $q\ne 0$,
which we will call the FMD (directional ferromagnetic) phase.  In the
full model, this condensation of $q$ corresponds to the condensation of
the vector field $A_\mu$, and $m^2$ corresponds to the gauge field mass
[\cite{golsha}].
The critical point $\kappa=\kappa_c$ ($=0$ at tree level),
$\tk\to\infty$ or $g\to 0$ should therefore correspond to the desired 
continuum limit,
with the desired chiral fermions and massless gluons, in perturbation
theory [\cite{sha}].

The discussion of the order parameter $\langle q_\mu\rangle$, however,
does not complete our discussion of the phase transition at
$\kappa=\kappa_c$.  Let us consider the expectation value
$v=\langle\phi_x\rangle$ for $\kappa>0$, where the tree-level scalar
propagator is given by the expression in \eq{props}.  To leading
order in $1/\tk$ we obtain
$$\eqalignno{
\langle\phi_x\rangle&=1-{1\over{4\tk}}\langle\theta_x^2\rangle+\dots\cr
&=1-{1\over{4\tk}}\int_p {1\over{\phat^2(\phat^2+m^2)}}+\dots,&(vev),}$$
where $\int_p=\int d^4p/(2\pi)^4$ is the integral over the Brillouin
zone.  For $m^2\to 0$ this is infrared divergent, and we need to resum
the series in order to obtain a finite answer:
$$\eqalignno{
\langle\phi_x\rangle&=\exp\left[-{1\over{4\tk}}\int_p G(p)\right]
\left(1+O\left({1\over{\tk^2}}\right)\right)\cr
&\sim (m^2)^{\eta}
\left(1+O\left({1\over{\tk^2}}\right)\right),&(vevresum)}$$
with 
$$\eta\equiv {1\over{64\pi^2\tk}}+O\left({1\over{\tk^2}}\right).
\eqno(eta)$$
The $O\left({1\over{\tk^2}}\right)$ corrections come from $\theta$
self-interactions, which we will discuss below.  We see that for
$\kappa\searrow\kappa_c$, $v$ goes to zero with a $\tk$-dependent
critical exponent $\eta$.  This situation is very reminiscent of that
with massless scalars in two dimensions, \cf\ the Coleman/Mermin--Wagner
theorem [\cite{col}].  It is simply a consequence of the fact that the
scalar propagator goes like $1/(p^2)^2$ for $m^2=0$.

\eq{vevresum} has a very important consequence: for $m^2\to 0$ ({\it i.e.}
$\kappa\to\kappa_c$), $\langle\phi_x\rangle$ goes to zero, and 
the full $U(1)_L\times U(1)_R$ symmetry ({\it cf.} 
\eqs{gaugesym,globalsym}) is restored at $\kappa=\kappa_c$.  
This implies that the $U(1)_L$ (and $U(1)_R$)
charges of massless fermions are well defined at the critical point. 

Interactions can be taken into account systematically in perturbation
theory. To order $1/\tk^2$, \eq{vevresum} is replaced by
$$\eqalignno{
\langle\phi_x\rangle&=\exp\left[-{1\over{4\tk}}\int_p G^{1-loop}(p)\right]
\cr&\sim (\kappa-\kappa_c^{1-loop})^{\eta},&(vevoneloop)}$$
where $G^{1-loop}$ differs from $G$ by finite wave function
and mass renormalizations.  Also the critical value of $\kappa$ is
shifted from its (vanishing) tree-level value to [\cite{bgs}]
$$\kappa_c^{1-loop}= 0.02993(1).\eqno(kappacrit)$$
The fact that the renormalizations are finite originates in the fact
that the interactions are irrelevant (in the abelian case), and
therefore do not change the long-distance behavior of correlation
functions.  See ref. [\cite{bgs}] for a much more detailed analysis of
the order parameter $\langle\phi_x\rangle$ in both the FM and FMD phases, 
where it is shown that perturbation theory agrees very well
with numerical results.
 
\subhead{\bf 4. Fermion spectrum in the reduced model}

In this section we will present one of the key results of this
paper: the fermion self-energy to one loop in the reduced model. But let
us first discuss what we would expect, if the reduced model is to pass
the test outlined in section 2.  The fermion action in \eq{Sreduced} is
formulated in terms of a charged left-handed field $\psi^c_L=P_L\psi$
(\ie\ it transforms under the symmetry \eq{gaugesym}), and a neutral
right-handed field $\psi^n_R=P_R\psi$ (which does not transform under
\eq{gaugesym}).  In the continuum limit, the neutral right-handed fermion
is free, because of the shift symmetry \eq{shiftsymm} [\cite{golpet}].
Moreover, at least naively, the charged left-handed fermion is also free
in the continuum limit, because the interaction terms in \eq{Sreduced}
with the field $\theta$ are irrelevant (in the usual technical sense,
{\it i.e.} dimension greater than four;
$\theta$ has mass dimension zero, {\it cf.} \eq{props}),
as can be seen by inserting and expanding \eq{theta}.  However, this
argument does not take into account the nonstandard infrared behavior
of the scalar field $\theta$, and might therefore be misleading.  We 
will therefore study the fermion propagator at one loop in perturbation
theory, and see that, to this order, the argument just given is 
nevertheless correct.  For a quicker, but more heuristic argument
leading to the same result, see ref. [\cite{lat97i}].

In order to perform actual perturbation theory calculations, it is
advantageous to reformulate the reduced action, \eq{Sreduced} by a field
redefinition of the fermion variables.  By redefining
$\psi^n_R=\phi^\dagger\psi^c_R$ or $\psi^c_L=\phi\psi^n_L$ we can write
the action in terms of respectively charged or neutral fermion fields
only.  This has the advantage of improving the infrared behavior of loop
corrections.  Here we will choose the charged option.  To order $1/\tk$, 
for $G=U(1)$, the reduced action becomes 
$$\eqalignno{
S_{\rm reduced}^{\rm fermion}=\half
\sum_{x,\mu}\Bigl\{&\psibar^c_x\gamma_\mu\psi^c_{x+\mu}
-\psibar^c_{x+\mu}\gamma_\mu\psi^c_x
-r\psibar^c_x(\Box\psi^c)_x\cr
&+{i\over{\sqrt{2\tk}}}(\partial^+_\mu\theta)_x
\;(\psibar^c_x\gamma_\mu P_R\psi^c_{x+\mu}+
\psibar^c_{x+\mu}\gamma_\mu P_R\psi^c_x)\cr
&-{1\over{4\tk}}(\partial^+_\mu\theta)_x^2
\;(\psibar^c_x\gamma_\mu P_R\psi^c_{x+\mu}-
\psibar^c_{x+\mu}\gamma_\mu P_R\psi^c_x)\cr
&-r\Bigl[{i\over{\sqrt{2\tk}}}(\partial^+_\mu\theta)_x
\;(\psibar^c_x\psi^c_{x+\mu}-\psibar^c_{x+\mu}\psi^c_x)\cr
&\phantom{-r\Bigl[}
-{1\over{4\tk}}(\partial^+_\mu\theta)_x^2
\;(\psibar^c_x\psi^c_{x+\mu}+\psibar^c_{x+\mu}\psi^c_x)\Bigr]\Bigr\},
&(Sredch)
}$$  
where $\partial^+_\mu$ is the forward derivative: $(\partial^+_\mu f)_x
=f_{x+\mu}-f_x$. If we would have chosen to use the neutral formulation,
the action would have been similar, but for a parity transformation
$P_L\leftrightarrow P_R$, $\theta\to -\theta$, and the omission of
scalar-fermion couplings proportional to $r$.  
Note that, in both formulations, the $\theta$ field always appears with
derivatives, improving the infrared behavior of perturbation theory in
the limit $m^2\to 0$.
(In the nonabelian case,
there would have been extra scalar-fermion couplings involving the
commutator $[\theta_x,\theta_{x+\mu}]$.  We believe that in this case
the infrared finiteness in the limit $m^2\to 0$ of observables invariant 
under the symmetries of
the model can be proven adapting the methods of ref. [\cite{david}].)    
The calculation of the charged
fermion one-loop self-energy proceeds in a straightforward manner.
There are two contributions, depicted in figure 1.  The tadpole diagram
of figure 1a gives a contribution
$$\Sigma^{(a)}(p)=
{1\over{8\tk}}\sum_\mu(-i\gamma_\mu\sin{p_\mu}P_R+r\cos{p_\mu})
\int_k\sum_\nu(1-\cos{k_\nu})G(k),\eqno(sigma2)$$
while the diagram of figure 1b leads to a more complicated contribution
$$\eqalignno{
\Sigma^{(b)}(p)=\;&{1\over{8\tk}}\sum_{\mu\nu}\e^{-ip_\mu+ip_\nu}
\int_k G(k)(\e^{-ik_\mu}-1)(\e^{ik_\nu}-1)\cr
\Bigl[&-\gamma_\mu i\slash s(k+p)\gamma_\nu P_R D^{-1}(k+p)
\;(\e^{i(k+2p)_\mu}+1)(\e^{-i(k+2p)_\nu}+1)\cr
&-rS(k+p)\gamma_\nu P_R\;(\e^{i(k+2p)_\mu}-1)(\e^{-i(k+2p)_\nu}+1)\cr
&-r\gamma_\mu\left(i\slash s(k+p)P_L-rM(k+p)P_R\right)D^{-1}(k+p)
\;(\e^{i(k+2p)_\mu}+1)(\e^{-i(k+2p)_\nu}-1)\cr
&-r^2 S(k+p)\;(\e^{i(k+2p)_\mu}-1)(\e^{-i(k+2p)_\nu}-1)\Bigr].
&(sigma1)}$$
The total one-loop self-energy is given by $\Sigma(p)=\Sigma^{(a)}(p)+
\Sigma^{(b)}(p)$.  

\vskip 1.0cm
%
  %
\epsfysize=4.0cm
\centerline{\epsfbox{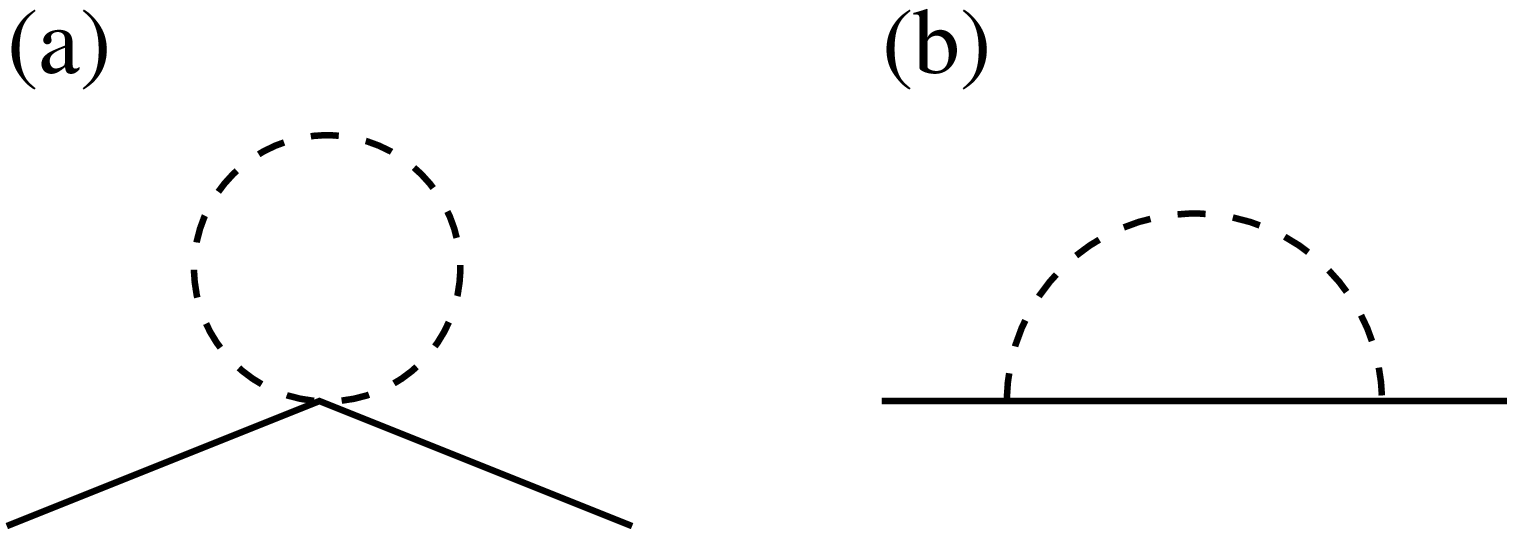} }
%
  %
\vskip 0.8cm
Fig. 1: One-loop fermion self-energy
\vskip 0.5cm

First, substituting $p=0$, we find $\Sigma(0)=0$, which
tells us that no mass counterterm is needed in order to keep the fermion
massless.  In the neutral formulation, this is a direct consequence of
shift symmetry [\cite{golpet}], and what we find here in the charged
formulation is consistent with that.

Next, we are interested in the nonanalytic behavior of the self-energy
in the continuum limit.  To start, let us see what happens to the
doublers, \ie\ for momenta $p=\pi_A+\tlp$ where we take $\tlp$ small and
$$\pi_A\in\{(\pi,0,0,0),\dots,(\pi,\pi,\pi,\pi)\}.\eqno(piAs)$$
The only pole in the fermion propagator in $\Sigma^{(b)}$ occurs for
$k=\pi_A+\tlk$ with $\tlk$ small, but in that region $G(k)$ is of order
one, and therefore these regions do not lead to any nonanalytic terms in
$\tlp$ in the continuum limit.  
For small $k$ of course $G^{-1}(k)\approx k^2(k^2+m^2)$, but
now $S(k+p)$ is of order one (thanks to the Wilson term), and again
there are no nonanalytic terms coming from this region.  (Note that the
derivative couplings of $\theta$ play an important role here!)  We conclude
that, for these momenta, $\Sigma(p)$ constitutes a small 
regular correction of order $1/\tk$, 
and that therefore the doublers are still removed by
the tree-level Wilson term.

For $p$ small (\ie\ $\pi_A=0$) all nonanalytic behavior comes from the
region around $k=0$. We obtain the nonanalytic terms by cutting out
a small region with radius $\delta$ around $k=0$, with $k\ll\delta\ll
1$, so that we can replace the integrand inside this region by its
covariant (continuum) expression [\cite{karsmit}]. 
(Any explicit $\delta$ dependence
coming from the region $k<\delta$ must cancel against the explicit 
$\delta$ dependence coming from the region $k>\delta$,
leaving the complete result independent of the arbitrary parameter $\delta$.)
Power counting tells us that no
contribution comes from any of the terms proportional to a power of $r$,
and we find, in the continuum limit,
$$\eqalignno{
\Sigma_{\rm nonan}(p)&={{-i}\over{2\tk}}\int_{|k|<\delta}
G(k)\slash k(\slash k+\slash p)\slash k P_R(k+p)^{-2}\cr
&={{-i\slash pP_R}\over{32\tk\pi^2}}\left(\log{{p^2}\over{\delta^2}}
+\half\left[\left({{p^2}\over{m^2}}+{{m^2}\over{p^2}}+2\right)
\log{\left(1+{{m^2}\over{p^2}}\right)}-{{m^2}\over{p^2}}
\log{{m^2}\over{p^2}}-1\right]\right)\cr
&\to {{-i\slash pP_R}\over{32\tk\pi^2}}\log{{p^2}\over{\delta^2}},
\ \ \ \ \ m^2\to 0,&(sigmanonan)
}$$
for small $p^2/\delta^2$.  This result shows that nonanalytic terms
occur only in the right-handed kinetic part of the charged
fermion propagator.  The left-handed kinetic term receives
only a finite renormalization coming from contact terms in the fermion
self-energy.   This tells us that the left-handed charged fermion is a
free particle, with a simple pole in its two-point function.

A similar analysis of the neutral propagator at one loop can be performed
by expressing \eq{Sreduced} in terms of the neutral fermion field
$\psi^n=\phi^\dagger\psi^c$.  One finds similar nonanalytic terms only in 
the left-handed kinetic part of the neutral fermion propagator, telling us
that in this case, the right-handed neutral fermion is free.  The finite 
one-loop
renormalization of the right-handed kinetic term actually vanishes in this
case, in accordance with shift symmetry.

If indeed the neutral right-handed fermion and the charged left-handed
fermion are the
only free fermions that exist at the critical point $m^2=0$
in the reduced model, one would
expect that the two-point functions of $\psi^c_R$ and $\psi^n_L$
correspond to two-point functions of fermion-scalar composite operators,
with a cut starting at $p=0$ (for $m^2=0$).  In fact, in the continuum
limit, we would expect to find that these correlation functions
factorize:
$$\langle\psi^c_{Rx}\psibar^c_{Ry}\rangle\sim
\langle\psi^n_{Rx}\psibar^n_{Ry}\rangle\langle\phi^\dagger_x\phi_y\rangle,
\eqno(factorize)$$
and similar for the neutral left-handed fermion.  We will show now
that the nonanalytic behavior found for the charged
right-handed fermion is exactly what one would obtain from calculating
the right-hand side of \eq{factorize} in momentum space, expanded to
order $1/\tk$.  An analogous argument can be given for the neutral
left-handed fermion.

The bosonic two-point function in \eq{factorize} is
$$\eqalignno{
\langle\phi^\dagger_x\phi_y\rangle&=\exp{1\over{2\tk}}[G(x-y)-G(0)]
\left(1+O\left({1\over{\tk^2}}\right)\right)\cr
&=1+{1\over{2\tk}}[G(x-y)-G(0)]+\dots,&(phicorr)
}$$
where
$$G(x-y)=\int_p\e^{i(x-y)}G(p).\eqno(Gxy)$$
In order to calculate $\langle\psi^n_{Rx}\psibar^n_{Ry}\rangle$, we need
to repeat the self-energy calculation, but now with $S_{\rm reduced}$ in
the neutral fermion formulation.  This calculation is analogous, but
simpler than the one we outlined above, so we will not repeat it here,
but just quote the results as we need them.  One finds that in this
case, the only nonanalytic term occurs for the left-handed fermion,
\ie\ the nonanalytic neutral self-energy is the parity-transformed
version of \eq{sigmanonan}.  We have
$$\eqalignno{
\sum_{xy}\e^{-ipx+iqy}\langle\psi^n_{Rx}&\psibar^n_{Ry}\rangle
\langle\phi^\dagger_x\phi_y\rangle=&(compfion)\cr
\delta(p-q)\;&\exp[-G(0)/2\tk]
\left[P_R S^n(p)P_L+
{1\over{2\tk}}\int_k P_R S^n(p-k)P_L G(k)+O\left({1\over{\tk^2}}\right)\right],
}$$ 
where $S^n$ is the neutral fermion propagator, and we wish
to calculate the right-hand side of \eq{compfion} to order $1/\tk$.
The first term in square brackets does not contain any nonanalytic terms
in the continuum limit, because of the chiral projectors.  The second
term, in which we may replace $S^n(p-k)$ by $S(p-k)$ to the desired
accuracy, yields the following nonanalytic terms in the continuum limit:
$$\eqalignno{
{1\over{2\tk}}&\int_k P_R S^n(p-k)P_L G(k)\to&(compnonan)\cr
&{1\over{32\pi^2\tk}}\left({{-i\slash p}\over{p^2}}\right)
i\slash p P_R\Biggl[\log{{p^2}\over{m^2}}\cr
&\phantom{{1\over{32\pi^2\tk}}\left({{-i\slash p}\over{p^2}}\right)
i\slash p P_R\Biggl[}
+\half\left(\left({{p^2}\over{m^2}}+{{m^2}\over{p^2}}+2\right)
\log{\left(1+{{m^2}\over{p^2}}\right)}-{{m^2}\over{p^2}}
\log{{m^2}\over{p^2}}-1\right)\Biggr]
\left({{-i\slash p}\over{p^2}}\right).
}$$
If we amputate the two massless fermion propagators, this expression is
not quite equal to minus the self-energy given in \eq{sigmanonan} yet.
For this, we need to include the nonanalytic part coming from expanding
the factor $\exp[-G(0)/2\tk]$ with
$$\eqalignno{
{{G(0)}\over{2\tk}}&={1\over{2\tk}}\int_p G(p)\cr
&=-\;{1\over{32\pi^2\tk}}\log{{m^2}\over{\delta^2}}+{\rm constant},
&(Gn)}$$
where we again isolated the nonanalytic term by cutting out a spherical
region with radius $\delta$ from the integration region.  Combining this
with the tree-level part of \eq{compfion} and with \eq{compnonan} we
recover exactly the expression \eq{sigmanonan} for the charged
right-handed fermion self-energy.  

The dynamics of the scalar field $\phi$ plays a crucial role in
obtaining this state of affairs.  A very similar model, the Smit--Swift
model [\cite{smiswi}], has been studied in the past with hopes of
enforcing the situation described above.  Without gauge fields, the
Smit--Swift model corresponds to \eq{Sreduced} with $\tk=0$.  For no
values of $\kappa$ and the Wilson--Yukawa coupling $r$ does one obtain
the desired result: if the global $G_L\times G_R$ symmetry is unbroken, 
neutral or charged massless fermions always come in left- and
right-handed pairs (for a review see ref. [\cite{petcher}]).  This is
in accordance with a general argument about the applicability of the
Nielsen--Ninomiya theorem 
[\cite{nienim}] to interacting theories [\cite{shann}]. 
Here we see that addition of an extra parameter, $\tk$, which has 
its origin in gauge fixing, makes it possible to construct a 
continuum limit in which the symmetry is unbroken, and the chiral
fermions undoubled.  For a discussion as to how this is not in
contradiction with the Nielsen--Ninomiya theorem, see ref.
[\cite{lat97ii}].

We will end this section with some remarks.
First, the calculation of the neutral and charged fermion propagators
could have been done starting directly from \eq{Sreduced}, in what we
will call the ``mixed formulation."  For the two-point functions
which are invariant
under the global symmetry, $\langle\psi^c_{Rx}\psibar^c_{Ry}\rangle$,
$\langle\psi^c_{Lx}\psibar^c_{Ly}\rangle$,
$\langle\psi^n_{Rx}\psibar^n_{Ry}\rangle$
and $\langle\psi^n_{Lx}\psibar^n_{Ly}\rangle$, we would have found 
exactly the same results.  (For noninvariant quantities, resummations are
necessary in order to remove infrared divergences; the simplest 
example of this is $\langle\phi_x\rangle$ discussed in the previous
section.)  This holds only for the connected correlation functions, and
not for ``auxiliary" quantities such as the self-energy. 

Second, we believe that all these arguments can be extended to higher 
orders in perturbation theory.  This is based on the observation that
the infrared structure of the reduced model is very similar to that
of two-dimensional theories with massless scalars.  There is a 
vast literature on such two-dimensional models, see {\it e.g.} refs.
[\cite{wit,david}], and we expect that some of the arguments and
methods can be adapted to our four-dimensional case. 

Last, we note that all arguments in this section
generalize to the nonabelian case.  

\subhead{\bf 5. Vacuum polarization}

Let us first consider the effects of the fermions on the dynamics of the 
scalar field, $\theta$. Since in the continuum limit the
gauge degrees of freedom, which are represented by the field $\theta$, are
supposed to decouple (after suitable adjustment of local counterterms), we
expect the lattice dynamics to conform with this expectation.  In particular,
we expect that no nonanalytic terms survive in the continuum limit of the
$\theta$ two-point function which come from fermion loops.  We will verify this
explicitly at the one-loop level.

It is convenient to perform this calculation using the neutral-fermion
language.  Of course, one would obtain the same
result using the charged-fermion form of $S_{\rm reduced}$
(\eq{Sredch}).  Expanded to order
$1/\tk$, the reduced action with neutral fermions is
$$\eqalignno{
S_{\rm reduced}^{\rm fermion}=\half
\sum_{x,\mu}\Bigl\{&\psibar^n_x\gamma_\mu\psi^n_{x+\mu}
-\psibar^n_{x+\mu}\gamma_\mu\psi^n_x-r\psibar^n_x(\Box\psi^n)_x\cr
&-{i\over{\sqrt{2\tk}}}(\partial^+_\mu\theta)_x
\;(\psibar^n_x\gamma_\mu P_L\psi^n_{x+\mu}+
\psibar^n_{x+\mu}\gamma_\mu P_R\psi^n_x)\cr
&-{1\over{4\tk}}(\partial^+_\mu\theta)_x^2
\;(\psibar^n_x\gamma_\mu P_L\psi^n_{x+\mu}-
\psibar^n_{x+\mu}\gamma_\mu P_L\psi^n_x)\Bigr\}.
&(Sredn)
}$$  
We define the $\theta$ self-energy $\Sigma_\theta$ from the full $\theta$
two-point function $G_{\rm full}$ by
$$G_{\rm full}^{-1}(p)=\phat^2(\phat^2+m^2)+\Sigma_\theta(p).\eqno(Sigmath)$$
To one loop, the fermionic contribution to $\Sigma_\theta(p)$ is
$\Sigma^{\rm fermion}_\theta(p)=\Sigma^{(a)}_\theta(p)+\Sigma^{(b)}_\theta(p)$
(\cf\ figure 2), with
$$\eqalignno{
\Sigma^{(a)}_\theta(p)&={1\over{2\tk}}\int_k\Biggl[
\sum_{\mu\nu}8\sin{\half p_\mu}\sin{\half p_\nu}\cos{(k_\mu-\half p_\mu)}
\cos{(k_\nu-\half p_\nu)}\cr
&\phantom{={1\over{2\tk}}\int_k\Biggl[}
\times\Biggl(\sin{k_\mu}\sin{(k_\nu-p_\nu)}+\sin{k_\nu}\sin{(k_\mu-p_\mu)}\cr
&\phantom{={1\over{2\tk}}\int_k\Biggl[\times\Biggl(\sin{k_\mu}}
-\delta_{\mu\nu}\sum_\lambda\sin{k_\lambda}\sin{(k_\lambda-p_\lambda)}\Biggr)
D^{-1}(k)D^{-1}(k-p)\Biggr],\cr
\Sigma^{(b)}_\theta(p)&={1\over{2\tk}}\int_k
\sum_\mu 8\sin^2{\half p_\mu}\sin^2{k_\mu}D^{-1}(k),&(Stheta)
}$$
where $D$ is given in \eq{props}.
In order to find the continuum limit of this
expression, we need to expand it to order $p^4$ (\cf\ \eq{Sigmath}).  First,
the order $p^2$ term is
$$\eqalignno{
{1\over{2\tk}}p^2\int_k\Biggl[&\left(\sum_\mu\sin^2{k_\mu}\cos^2{k_\mu}-\half
\sum_\mu\sin^2{k_\mu}\sum_\nu\cos^2{k_\nu}\right)D^{-2}(k)\cr
&+\half\sum_\mu\sin^2{k_\mu}D^{-1}(k)\Biggr]
\approx 0.05464\times{1\over{2\tk}}p^2 
\ \ \ ({\rm for\ } r=1),&(psq)
}$$
leading to a one-loop contribution to $\kappa_c$ (\cf\ section 3)
$$\kappa_c^{1-loop}=0.02993(1)-0.02732(1)n_f\ \ \ ({\rm for\ } r=1),\eqno(kcf)$$
where $n_f$ is the number of left-handed fermions in the abelian
case.  

\vskip 1.0cm
%
  %
\epsfysize=4.0cm
\centerline{ \epsfbox{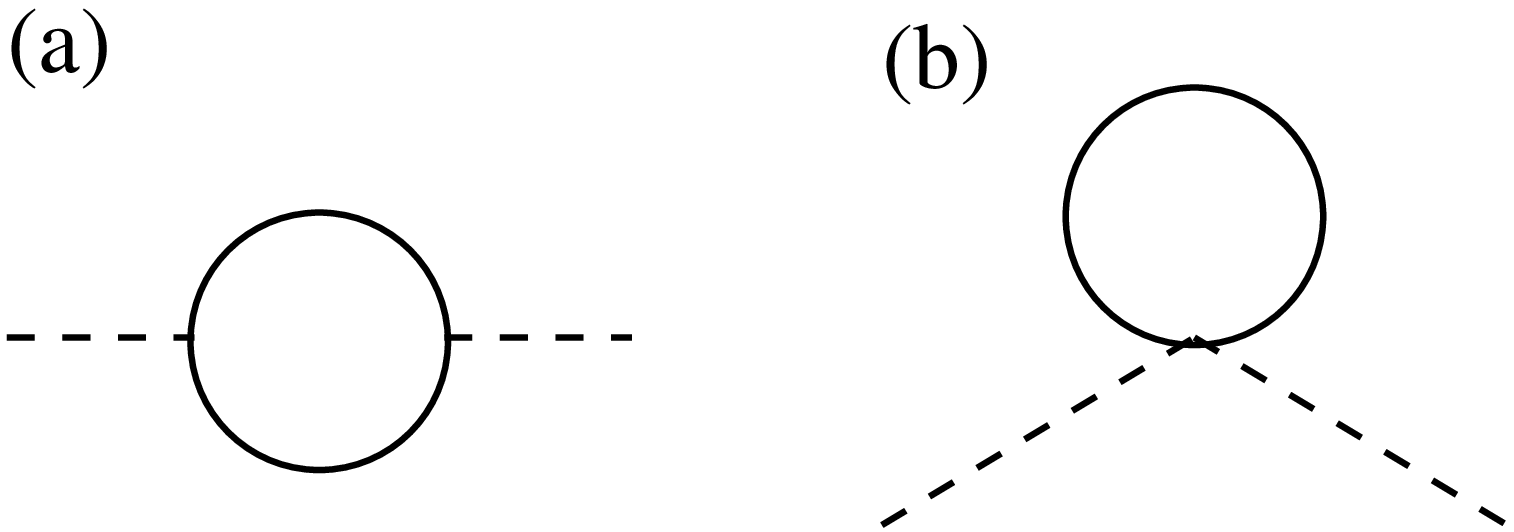} }
%
  %
\vskip 0.8cm
Fig. 2: One-loop $\theta$ self-energy
\vskip 0.5cm

Next, we are interested in the $O(p^4)$ term.  We will not calculate the
complete coefficient of this term, but restrict ourselves to inspection of the
nonanalytic part.  Like before, we can do this by restricting the
loop-momentum integration to the region $|k|<\delta$, and replacing the
integrand by its covariant form:
$$\Sigma_\theta(p)\sim{1\over{2\tk}}\sum_{\mu\nu}p_\mu p_\nu
I_{\mu\nu}(p),\eqno(Sthnonan)$$ 
where
$$\eqalignno{
I_{\mu\nu}(p)&=2\int_{|k|<\delta}{{2k_\mu k_\nu-k_\mu p_\nu-k_\nu p_\mu
-\delta_{\mu\nu}(k^2-k\cdot p)}\over{k^2(k-p)^2}}\cr
&={1\over{24\pi^2}}(p_\mu p_\nu-\delta_{\mu\nu}p^2)\log{p^2/\delta^2}
+{\rm regular\ terms}
&(I)
}$$
for $p^2\ll\delta^2$.  We see that, because the nonanalytic part of 
$I_{\mu\nu}$ is transversal, there
is indeed no nonanalytic contribution to the $\theta$ two-point function from
the fermions.  This again demonstrates that the reduced model leads to a theory
of free chiral
fermions decoupled from the gauge degrees of freedom in the continuum
limit, after a suitable tuning of local counterterms (in this case the
$\kappa$-term).

It is straightforward to verify that, in the abelian case, the same conclusion
holds for the one-loop contribution from the $\theta$ self-interactions, in
accordance with the fact that these self-interactions correspond to irrelevant
operators. 

Next, we would like to discuss the effective action for the gauge field
$A_{x,\mu}$, obtained by integrating out all other degrees of freedom.
An important test of our approach consists of the following. 
Take the external gauge field to be smooth.  The effective action can now
be defined in two ways:

\item{1.} We may set $\phi_x=I$ ({\it cf.} \eqs{SV,VHrel}), and integrate over 
the fermions.
\item{2.} We may integrate over both $\phi_x$ and the fermions
({\it cf.} \eq{SH}).

Both methods (which, in the terminology of section 2, correspond to 
respectively the ``vector" and ``Higgs" picture)
should yield the {\it same} gauge invariant effective action in
the continuum limit, modulo local counterterms
(if the fermion representation is anomaly-free).  The second
method verifies that the integration over the (lattice)
gauge orbit of the external
gauge field does not change the (long-distance part of the) effective action.

\vskip 0.8cm
%
  %
\epsfysize=4.0cm
\centerline{ \epsfbox{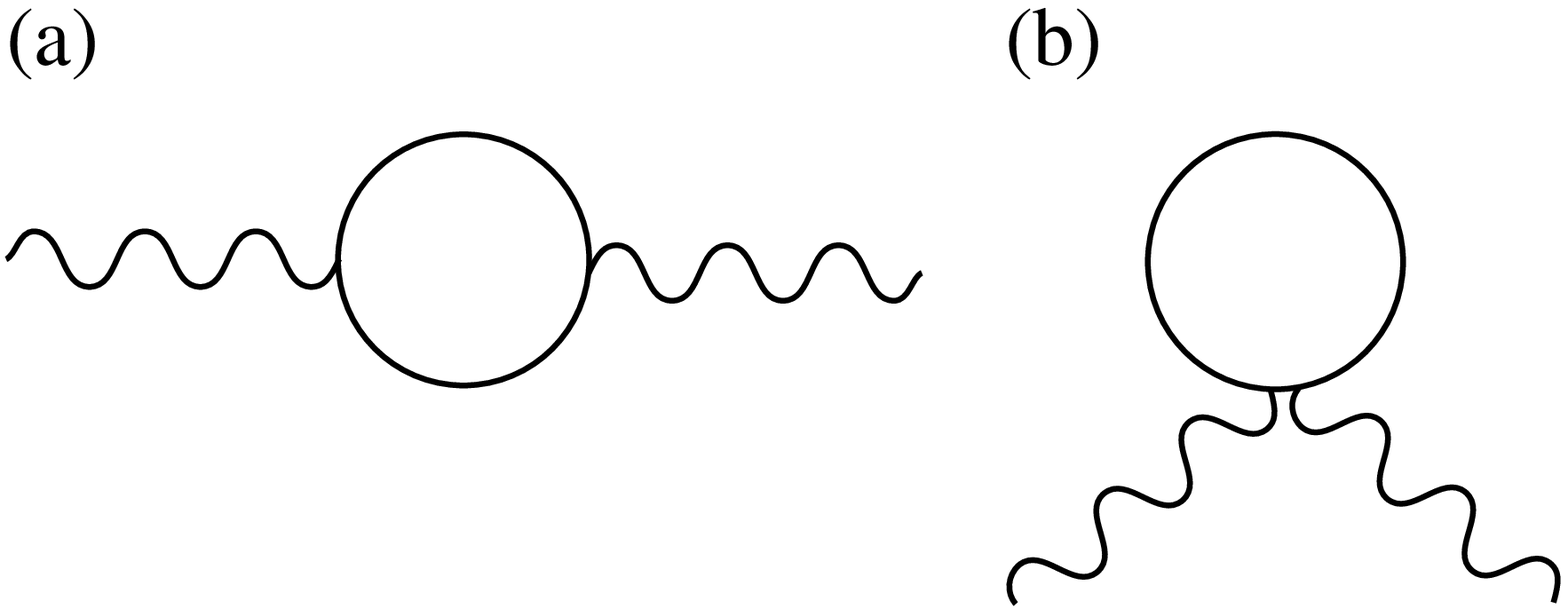} }
%
  %
\vskip 0.8cm
Fig. 3: One-loop contributions to the vacuum polarization 
\vskip 0.5cm

We will now examine this using the example of the fermionic contribution
to the abelian vacuum
polarization $\Pi_{\mu\nu}(p)$.  Starting from \eq{SV} 
(\ie\ following method 1), the vacuum polarization is just the sum
of the two one-loop diagrams of figure 3. We find that
$\Pi_{\mu\nu}(p)$ is given by the expression \eq{Sthnonan} for the
$\theta$ self-energy with a factor $p_\mu p_\nu/(2\tk)$ omitted. 
This leads to a one-loop gauge-field
mass counterterm in \eq{Sct} with $\kappa$ given by \eq{kcf}. For the
nonanalytic part, we find $\Pi_{\mu\nu}(p)=I_{\mu\nu}(p)$
(\eqs{Sthnonan,I}), leading to the one-loop $\beta$-function
$$\beta(g)\equiv{{\partial g}\over{\partial\log{a}}}=
-n_f{{g^3}\over{24\pi^2}}\eqno(betaf)$$
(in the nonabelian case this has to be multiplied by the appropriate quadratic
Casimir).  This is exactly the result we expect for ``chiral QED" with $n_f$
left-handed fermions.

Next, we wish to verify that the orbit integration does not change the
nonlocal part of the vacuum polarization (\cf\ method 2).
To one loop, this is trivial, since, as before, the vacuum polarization
is just the sum of the two diagrams of figure 3, which do not contain any 
$\theta$-lines.  Diagrams with internal $\theta$-lines only show up at two
loops and higher, and we did not perform an explicit calculation of these
diagrams.  But, one can easily understand on general grounds that
these higher-loop diagrams with internal $\theta$-lines do not contribute to
the nonanalytic part of the vacuum polarization, and that therefore their
effects can be removed by counterterms. 
Two- and higher-loop contributions can be conveniently calculated by
rewriting the action \eq{SH} in terms of charged fermion fields only, by
making the substitution $\psi=(P_L+P_R\phi^\dagger)\psi^c$ in \eq{SH}.
As in the reduced model, this improves the infrared behavior of 
perturbation theory, validating standard power-counting arguments in
particular.  We observe that, in the charged-fermion language,
gauge field-fermion vertices only occur in the left-handed kinetic term,
while $\theta$-fermion vertices occur only in the right-handed kinetic
and Wilson terms. (The reduced model in the charged-fermion formulation
is then obtained again by setting $U_{x,\mu}=1$.) 

\vskip 1.0cm
%
  %
\epsfxsize=14.0cm
\centerline{ \epsfbox{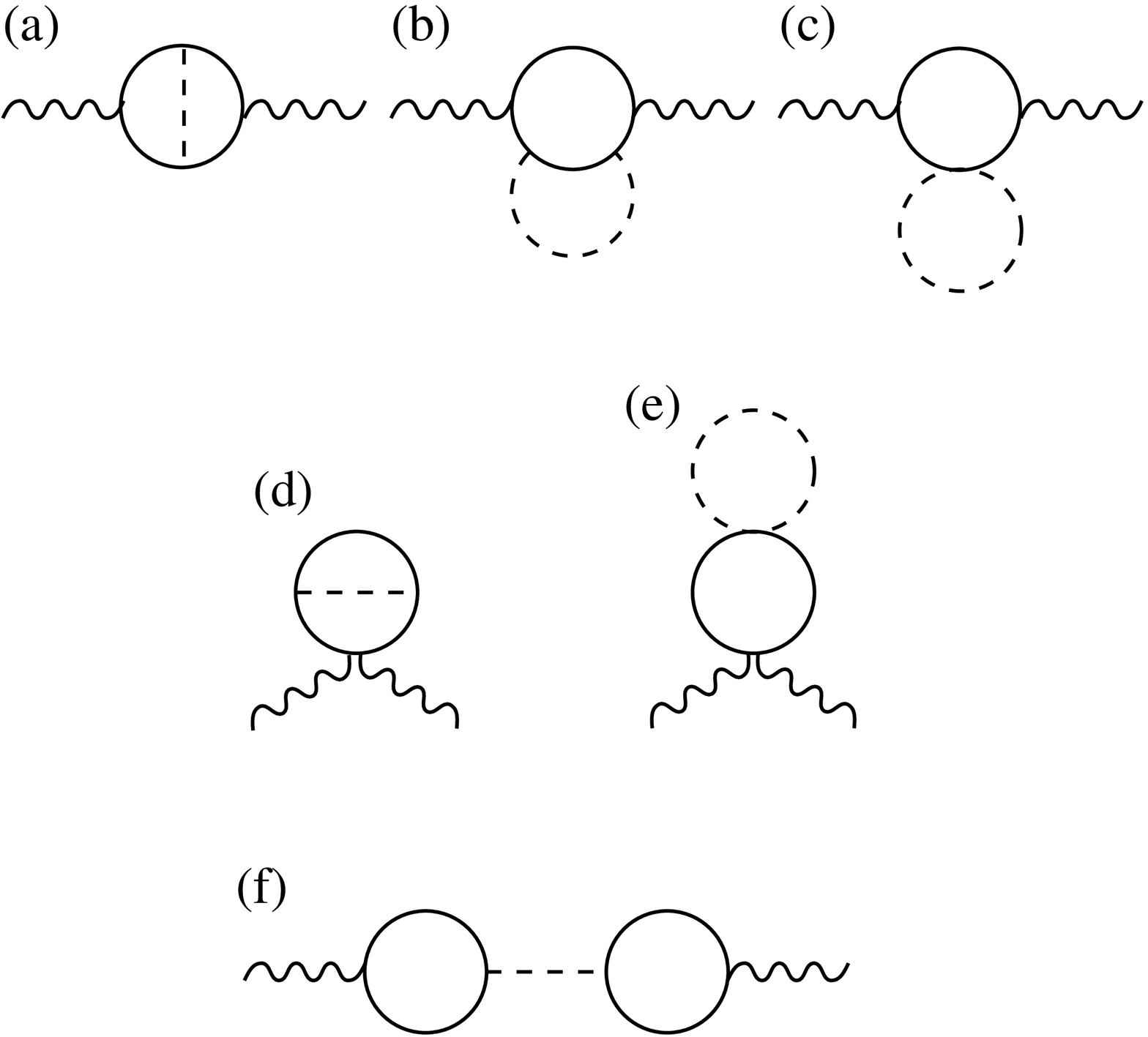} }
%
  %
\vskip 0.8cm
Fig. 4: Two-loop contributions to the vacuum polarization 
\vskip 0.5cm

The topology of the contributing two-loop diagrams is shown in
figure 4 (where we omitted any diagrams with $\theta$-tadpoles, since
this tadpole vanishes).  As we just explained,
fermion-$\theta$ vertices either arise from the Wilson term, or contain a
factor $\gamma_\mu P_R$.  If we start following the fermion
loop from one of the $A_\mu$ vertices, we
either encounter a vertex from the Wilson term, which corresponds to an
irrelevant operator, or we encounter a vertex $\gamma_\mu P_R$.  In this case,
because of the left-handed projector $P_L$ at the $A_\mu$ vertex, only the
Wilson term part of the fermion propagator (\cf\ \eq{props}) contributes,
which again corresponds to an irrelevant operator.  In both cases, we therefore
do expect these diagrams to yield only contact terms in the continuum limit. 

This analysis demonstrates explicitly how the ``rough gauge field problem"
is resolved within the gauge-fixing approach [\cite{sha}] also for nontrivial
orbits.  The resolution is a direct consequence of the fact that the
full theory, including the gauge degrees of freedom, can be systematically
investigated in perturbation theory.  

\subhead{\bf 6. The fermion-number current}

The fermion action, \eq{Sfermion}, is invariant under simple $U(1)$ phase
rotations of the fermion field
$$\psi\to e^{i\alpha}\psi\;,\ \ \ \ \psibar\to\psibar
e^{-i\alpha}\;.\eqno(fermionnumber)$$
This exact symmetry appears to be problematic, since it seems to imply
that we can define a continuum limit containing only left-handed
fermions (the right-handed fermions decouple in the continuum limit)
with a conserved U(1) quantum number [\cite{banks}].  
This would be in contradiction
with the fact that this $U(1)$ quantum number should be anomalous,
leading to fermion number violating processes through instantons
[\cite{thooft}]. Here, we analyze this question in 
perturbation theory, leaving a discussion of nonperturbative issues
to future work.  In this section, we will work in the vector picture,
\cf\ \eq{SV}.

The conserved current corresponding to the symmetry \eq{fermionnumber}
is
$$\eqalignno{
J_{x,\mu}&=J^L_{x,\mu}+J^R_{x,\mu}+J^W_{x,\mu}\;,\cr
J^L_{x,\mu}&={1\over 2}\left(\psibar_x\gamma_\mu
P_LU_{x,\mu}\psi_{x+\mu}+\psibar_{x+\mu}\gamma_\mu P_LU^\dagger_{x,\mu}
\psi_x\right)\;,\cr
J^R_{x,\mu}&={1\over 2}\left(\psibar_x\gamma_\mu
P_R\psi_{x+\mu}+\psibar_{x+\mu}\gamma_\mu P_R
\psi_x\right)\;,\cr
J^W_{x,\mu}&=-{r\over 2}\left(\psibar_x\psi_{x+\mu}
-\psibar_{x+\mu}\psi_x\right)\;.&(currents)
}$$
On the lattice, we have
$$\sum_\mu\left(J_{x,\mu}-J_{x-\mu,\mu}\right)=0\;.\eqno(cons)$$
However, $J_{x,\mu}$ is not 
gauge invariant, and therefore will not correspond to the appropriate
physical current in the continuum limit.  Let us consider this in
some detail by calculating the expectation value of the current,
$\langle J_{x,\mu}\rangle_A$ to quadratic order in the gauge
fields $A_{x,\mu}$, in the continuum limit. ($\langle\cdots\rangle_A$
denotes the functional average over $\psi$ and $\psibar$ only.)
We choose the fermions to be in the fundamental representation of
the gauge group $G$, and we write
$$A_{x,\mu}=A^a_{x,\mu}T_a\;,
\ \ \ \ A^a_{x,\mu}=\int_p e^{ipx}A^a_\mu(p)\;, \eqno(generators)$$
with $T_a$ the hermitian generators of the group $G$, normalized by
$$\tr T_a T_b=\half\;\delta_{ab}\;.\eqno(normal)$$

\vskip 1.0cm
%
  %
\epsfysize=4.0cm
\centerline{ \epsfbox{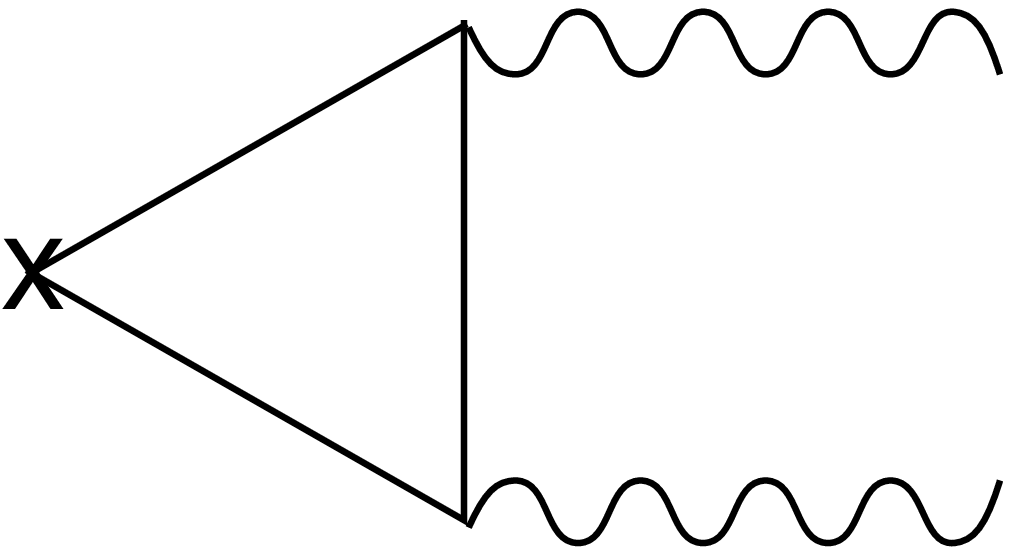} }
%
  %
\vskip 0.8cm
Fig. 5: Triangle diagram (the cross denotes the conserved current of
\eq{currents})            
\vskip 0.5cm

The only diagram that contributes to order $A^2$ is shown in figure 5.
(All other ``lattice artifact" diagrams vanish, as already observed in 
ref. [\cite{karsmit}].)  The parity-even part vanishes, and we find 
the following result, to
leading order in the gauge-field momenta $k$ and $l$:
$$\eqalignno{
\langle J^L_{x,\mu}\rangle_A&=
i\int_{kl} e^{i(k+l)x}[I_{\mu\rho\sigma}(k,l)+
I_L\;\epsilon_{\mu\nu\rho\sigma}(k-l)_\nu]
A^a_\rho(k)A^a_\sigma(l)\;,&(expvalues)\cr
\langle J^R_{x,\mu}\rangle_A&=
i\int_{kl} e^{i(k+l)x}I_R\;\epsilon_{\mu\nu\rho\sigma}(k-l)_\nu\;
A^a_\rho(k)A^a_\sigma(l)\;,\cr
\langle J^W_{x,\mu}\rangle_A&=
i\int_{kl} e^{i(k+l)x}I_W\;\epsilon_{\mu\nu\rho\sigma}(k-l)_\nu\;
A^a_\rho(k)A^a_\sigma(l)\;,
}$$
with summation implied over repeated indices. The function
$I_{\mu\rho\sigma}(k,l)$ is given by
$$\eqalignno{
I_{\mu\rho\sigma}(k,l)=
&\;2\epsilon_{\alpha\beta\mu\sigma}k_\alpha l_\beta
[k_\rho(I_{20}-I_{10})-l_\rho I_{11}]
+2\epsilon_{\alpha\beta\mu\rho}k_\alpha l_\beta
[k_\sigma I_{11}-l_\sigma(I_{02}-I_{01})]\cr
&+\half\epsilon_{\alpha\mu\rho\sigma}\Bigl(
k_\alpha[k^2I_{20}-2k\!\cdot\!lI_{11}+l^2(I_{02}-2I_{01})]\cr
&\phantom{+\half\epsilon_{\alpha\mu\rho\sigma}}
-l_\alpha[l^2I_{02}-2k\!\cdot\!lI_{11}+k^2(I_{20}-2I_{10})]\Bigr)
\;,&(inner)
}$$ 
where
$$I_{st}\equiv I_{st}(k,l)={1\over{16\pi^2}}\int_0^1 dx \int_0^{1-x}dy
{{x^sy^t}\over{
x(1-x)k^2+y(1-y)l^2+2xyk\!\cdot\! l}}\;.\eqno(Ist)$$
The constants $I_L$, $I_R$ and $I_W$ are given by
$$\eqalignno{
I_L&=r^2\int_p\left({1\over 4}c_1c_2c_3c_4
s^2(p)M^2(p)-c_1c_2c_3s^2_4s^2(p)M(p)
\right)D^{-4}(p)\;,&(outer)\cr
I_R&=r^2\int_p\left(\half c_1c_2c_3c_4
M^2(p)D^{-3}(p)-{1\over 4}[c_1c_2c_3c_4
s^2(p)+4r^2c_1c_2c_3s^2_4M(p)]M^2(p)
D^{-4}(p)\right)\;,\cr
I_W&=-r^2\int_p c_1c_2c_3s^2_4M(p)D^{-3}(p)\;,
}$$
in which
$$
s_\mu\equiv\sin{p_\mu}\;,\ \ \ c_\mu\equiv\cos{p_\mu}\;,
\ \ \ s^2(p)\equiv\sum_\mu\sin^2{p_\mu}\;.\eqno(defs)
$$
Lattice loop integrals were calculated again by splitting the
integration region into a small region with radius $\delta$ around
$p=0$ (``inner region"), and the rest (``outer region"), taking the
double limit $a\to 0$, followed by
$\delta\to 0$. (The split into inner and 
outer regions depends on the routing of the external momenta through
the loop: we chose the momentum of the fermion propagator connecting
the two gauge-field vertices to be $p-\half(k-l)$.  Of course, the
sum of inner and outer region contributions does not depend on this.) 
For $\langle J^R_{x,\mu}\rangle_A$ and $\langle J^W_{x,\mu}
\rangle_A$ the inner-region integrals vanish, while the integral
$I_{\mu\rho\sigma}(k,l)$ represents the (nonlocal)
inner-region contribution for $\langle
J^L_{x,\mu}\rangle_A$.  The integrals $I_{L,R,W}$ represent
outer-region contributions.  In other words, only $\langle
J^L_{x,\mu}\rangle_A$ is nonlocal, as one expects, since the
right-handed fermions are free in the continuum limit.
Using [\cite{karsmit}]
$$I_L+I_R+I_W=-{1\over{64\pi^2}}\eqno(ksint)$$
(for any nonzero value of $r$!), and
$$(k+l)_\mu I_{\mu\rho\sigma}(k,l)=
{1\over{64\pi^2}}\;\epsilon_{\mu\nu\rho\sigma}(k+l)_\mu(k-l)_\nu\;,$$
we find that indeed $\partial_\mu\langle J_{x,\mu}\rangle_A$ 
vanishes.

{}From \eqs{expvalues,inner,ksint} one can show that

$$k_\rho{{\delta\langle J_{x,\mu}\rangle_A}\over{\delta A^a_\rho(k)}}
={i\over{16\pi^2}}\int_l e^{i(k+l)x}\epsilon_{\mu\nu\rho\sigma}
k_\rho l_\nu A^a_\sigma(l)\;.\eqno(gaugeanom)$$
(In deriving this result, we used the relation $k^2(I_{10}(k,l)
-2I_{20}(k,l))=l^2(I_{01}(k,l)-2I_{02}(k,l))$.)  This proves that, as 
expected, the current $J_{x,\mu}$ is indeed
not gauge invariant, as was pointed out
in this context in ref. [\cite{dugman}].  A gauge invariant vector
current can be defined by adding an irrelevant term $J^{\rm irr}_{x,\mu}$
to $J_{x,\mu}$, with an expectation value that
goes to $K_\mu$ in the continuum limit, where [\cite{dugman}]
$$\langle J^{\rm irr}_{x,\mu}\rangle_A\to 
K_{x,\mu}={1\over{16\pi^2}}\;\epsilon_{\mu\nu\rho\sigma}\;
\tr(A_{x,\nu}F_{x,\rho\sigma}-{1\over 3}A_{x,\nu}A_{x,\rho}A_{x,\sigma})\;.
\eqno(Kcurr)$$
For example, we may take
$$J^{\rm irr}_{x,\mu}={1\over{32\pi^2I_W}}\;J^W_{x,\mu}\;.\eqno(irrcurr)$$
The current $J_{x,\mu}+J^{\rm irr}_{x,\mu}$ yields the correct,
gauge invariant, fermion-number current in the continuum limit to 
order $A^2$. Its divergence is
$$\partial_\mu\langle J_{x,\mu}+J^{\rm irr}_{x,\mu}\rangle_A\to
\partial_\mu K_{x,\mu}={1\over{32\pi^2}}\;\epsilon_{\mu\nu\rho\sigma}\;
\tr(F_{x,\mu\nu}F_{x,\rho\sigma})\;.\eqno(divcurr)$$
(An additional irrelevant operator of order $A^3$ would likely
be needed in order to construct a gauge invariant current to order $A^3$
in the nonabelian case.) 
Note that the vector current that leads to gauge invariant correlation
functions in the continuum limit, is {\it not} what one might naively
guess: $J^L_{x,\mu}+J^R_{x,\mu}$.  The reason is that, although this
operator is invariant under gauge transformations, the Feynman rules of
the theory are not.

\subhead{\bf 7. Conclusion and discussion}

In this paper, we studied a proposal for the construction of lattice
chiral gauge theories in (one-loop) weak-coupling perturbation theory.
We considered mostly the abelian case, and demonstrated that, in
perturbation theory, the model defined in section 2 has a continuum
limit with the desired chiral fermions, in which the gauge degrees of
freedom decouple, and with the correct one-loop $\beta$-function for the
gauge coupling. Note that, in the reduced model, the field $\theta$,
which represents the gauge degrees of freedom in the full model,
decouples from the fermions for any fermion content.  This is consistent
with the fact that the anomaly vanishes for a purely longitudinal
gauge field.  Together with the nonperturbative results presented
in refs. [\cite{bgslett,bgs}], this makes us confident that the
gauge-fixing approach can indeed be used to define abelian chiral gauge
theories on the lattice.  Of course, when the full dynamics of the gauge
field is taken into account, the fermion representation has to be
anomaly-free.  A next step (in the abelian case) would be to
investigate the potential between two static charges.  In principle,
the full counterterm action $S_{\rm ct}$ will have to be calculated, and
it would be interesting to see to what precision the counterterms 
have to be adjusted in order to obtain the Coulomb potential.
 
Our results should also apply to other lattice fermion
formulations, such as staggered fermions, domain wall fermions, or
Weyl fermions with Majorana mass and Wilson terms. 
(The latter were discussed in ref. [\cite{sha,rommaj}].
We verified explicitly that at one loop in the reduced model, the bare
Majorana mass can be tuned such that a free Weyl fermion emerges in the
continuum limit. Since in this case there is no shift symmetry, the
critical value of the bare fermion mass does not vanish.)
 
We expect that all perturbative results presented in this paper generalize
to the nonabelian case, with suitable modification.  For instance, the
long distance behavior of the gauge degrees of freedom (without fermions)
should be described by the continuum higher-derivative sigma model of
ref. [\cite{hata}], and we expect that it will. The analysis of the
fermion self-energy of section 4 carries over without change, and therefore
we expect the same conclusions about the fermion content as in
the abelian case.  The main reason that we have not considered the
nonabelian case in more detail here is that, in our view, nontrivial
nonperturbative issues will have to be addressed first.  The approach
to lattice chiral gauge theories investigated here
is inherently based on gauge-fixing. This raises the issue of Gribov
copies, which should be resolved before the proposal is ``complete" for
nonabelian gauge theories.  A related observation is that the BRST
approach to nonabelian gauge theories has not been defined outside
perturbation theory. Until this issue is better understood,
it is relatively less important to study the nonabelian
case in perturbation theory in much detail.  We note here that the fact
that our lattice gauge-fixing action $S_{\rm gf}$ has a unique global
minimum at $U_{x,\mu}=1$, while suppressing rough ``lattice" Gribov copies,
does not tell us anything about long-distance, continuum Gribov copies. 

Finally, we addressed the issue of fermion-number nonconservation, but only
in perturbation theory.  Work on the nonperturbative aspects of this issue
is in progress, and we expect to report on it in a future publication.
Here we just quote ref. [\cite{bhs}], in which it was shown that the
existence of a gauge-noninvariant conserved charge on the lattice
does not imply that fermion number is conserved.

\subhead{\bf Acknowledgements}

We would like to thank Sumit Das, Michael Ogilvie, Andrei Slavnov and
Pierre van Baal for discussions. WB is supported by the Deutsche
Forschungsgemeinschaft under grant Wo 389/3-2, MG by
the US Department of Energy as an Outstanding Junior Investigator,
and YS by the US-Israel Binational Science
Foundation and the Israel Academy of Science.

\references

\refis{bgslett}
W.~Bock, M.F.L.~Golterman and Y.~Shamir, hep-lat/9709154.

\refis{shalat95}
Y.~Shamir, \npb\ (Proc.\ Suppl.) {\bf 47} (1996) 212.

\refis{golsha}
M.F.L.~Golterman and Y.~Shamir, \plb{399} (1997) 148.

\refis{neu}
H.~Neuberger, \plb{183} (1987) 337;
B.~Sharpe, Jour. Math. Phys. {\bf 25} (1984) 3324.

\refis{smiswi}
P.D.V.~Swift, \plb{145} (1984) 256; J.~Smit, Acta Phys. Polonica
{\bf B17} (1986) 531. 

\refis{golpet}
M.F.L.~Golterman and D.N.~Petcher, \plb{225} (1989) 159.

\refis{rome}
A.~Borelli, L.~Maiani, G.-C.~Rossi, R.~Sisto and M.~Testa, \plb{221}
(1989) 360; \npb{333} (1990) 335; L.~Maiani, G.-C.~Rossi and
M.~Testa, \plb{292} (1992) 397.

\refis{foenie}
D.\ Foerster, H.B.\ Nielsen and M.\ Ninomiya,
\plb{94} (1980) 135; J.\ Smit, \npb\ (Proc.\ Suppl.) {\bf 4} (1988) 451;
S.\ Aoki, \prl{60} (1988) 2109;
K.\ Funakubo and T.\ Kashiwa, \prl{60} (1988) 2113.

\refis{sha}
Y.~Shamir, \prd{57} (1998) 132. 

\refis{bgs}
W.~Bock, M.F.L.~Golterman and Y.~Shamir, hep-lat/9708019. 

\refis{col}
S.~Coleman, \cmp{31} (1973) 259; N.D.~Mermin and H.~Wagner, 
\prl{17} (1966) 1133.

\refis{petcher}
D.N.~Petcher, \npb\ (Proc.\ Suppl.) {\bf 30} (1993) 50.

\refis{nienim}
H.~Nielsen and M.~Ninomiya, \npb{185} (1981) 20; Erratum
ibid. {\bf B195} (1982) 541; \npb{193} (1981) 173.

\refis{shann}
Y.~Shamir, \prl{71} (1993) 2691; \npb\ (Proc.\ Suppl.) {\bf 34}
(1994) 590; hep-lat/9307002.

\refis{david}
F.~David, \plb{96} (1980) 371; \cmp{81} (1981) 149.

\refis{karsmit}
L.~Karsten and J.~Smit, \npb{183} (1981) 103.

\refis{lat97i}
W.~Bock, M.F.L.~Golterman and Y.~Shamir, to be published in the
Proceedings of Lattice 97, the International Symposium on Lattice
Gauge Theory, Edinburgh, Scotland, hep-lat/9709113.

\refis{lat97ii}
W.~Bock, M.F.L.~Golterman and Y.~Shamir, to be published in the
Proceedings of Lattice 97, the International Symposium on Lattice 
Gauge Theory, Edinburgh, Scotland, hep-lat/9709115.

\refis{wit}
E.~Witten, \npb{145} (1978) 110.

\refis{banks}
T.~Banks, \plb{272} (1991) 75.

\refis{thooft}
G.~'t~Hooft, \prl{37} (1976) 8; \prd{14} (1976) 3432; Erratum ibid. {\bf
D18} (1978) 2199.

\refis{dugman}
M.J.~Dugan and A.V.~Manohar, \plb{265} (1991) 137.

\refis{hata}
H.~Hata, \plb{143} (1984) 171; see also Y.~Kikukawa, hep-lat/9707010.

\refis{bhs}
W.~Bock, J.E.~Hetrick and J.~Smit, \npb{437} (1995) 585.

\refis{rommaj}
L.~Maiani, G.C.~Rossi and M.~Testa, \plb{292} (1992) 397;
G.~Travaglini, \npb{507} (1997) 709.

\endreferences

\vfill
\bye